\documentclass{article}
\usepackage[a4paper, total={16cm, 24.7cm}]{geometry}

\usepackage{graphicx} 
\usepackage{float} 
\usepackage[utf8]{inputenc} 
\usepackage{url} 
\usepackage[square,numbers]{natbib} 
\usepackage{authblk} 

\newcommand{\tsup}[1]{\textsuperscript{#1}}
\newcommand{\tsub}[1]{\textsubscript{#1}}

\title{Advantages of off-line analysis of digitally recorded pulses in case of neutron-gamma discrimination in scintillators}
\author[1,2]{Lukasz Adamowski}
\author[1]{Martyna Grodzicka-Kobylka}
\author[1]{Tomasz Szczesniak}
\author[1]{Agnieszka Syntfeld-Ka\.zuch}
\author[1]{Lukasz Swiderski}
\author[2]{Adam Kisiel}
\affil[1]{National Centre for Nuclear Research (NCBJ), Otwock-Swierk, Poland}
\affil[2]{Faculty of Physics, Warsaw University of Technology, Warsaw, Poland}

\begin{document}

\maketitle

\begin{abstract}
Many modern digital analyzers offer the ability to record raw pulses from ionizing radiation detectors. We use this opportunity to investigate the effectiveness of Charge Comparison Method in Pulse Shape Discrimination of neutron and gamma radiation measured with organic glass scintillator and \mbox{trans-stilbene}. The idea of software for automated off-line analysis of digitally recorded data is briefly described.
We discuss the difference between Leading Edge and Constant Fraction Discrimination triggering methods and we propose triggering on pulse maximum as an alternative.
We observe that the starting point of charge integration gates has major impact on Figure of Merit values, therefore it is important to choose it carefully and report it with other Charge Comparison Method parameters to keep comparison between scintillators reliable.
Figure of Merit has a limited usage, so Relative Height of Minimum is proposed as an additional indicator of neutron-gamma discrimination effectiveness in practical applications.
\end{abstract}

\section{Introduction}

In recent years, there has been a growing trend in the development of detectors capable of measuring various types of ionizing radiation simultaneously. One example is organic scintillators that demonstrate the ability to discriminate between neutrons and gamma rays. This capability arises because hydrogen-rich scintillators enable fast neutrons to interact via recoil protons (hydrogen nuclei), producing pulses longer than those of gamma rays, which primarily interact with electrons.
\par
Pulse Shape Discrimination (PSD) is a widely used technique to distinguish neutrons from gamma rays, and the Charge Comparison Method (CCM) is a simple and effective implementation of this approach. Historically, CCM has been employed in analog detection systems and continues to be used in modern digital analyzers. 
Many more sophisticated PSD methods have been developed over the years.
Pulse Gradient Analysis \cite{DMellow2007} investigates the slope of pulses.
Frequency Gradient Analysis uses the Fourier transform to exploit the difference in frequency components and the Discrete Wavelet Transform adds a time factor to this analysis \cite{Singh2024}.
Time-over-threshold has been proven successful when an adaptive noise filter is applied \cite{Paul2024}.
Some methods compare pulses with previously measured reference shapes \cite{Ambers2011} and there is a vast family of Neural Networks of many kinds, that are trained with a large amount of data (measured or simulated) to distinguish neutrons from gamma rays \cite{Fabian2021, Liu2022, Tingmeng2024}.
They achieve results that are undoubtedly better than CCM and can perform tasks that CCM is not able to perform (such as pile-up rejection \cite{Fu2018}), but for the price of high computational power demand. Although computing is more affordable now thanks to the progress in FPGA integrated circuits and Machine Learning specific hardware, CCM remains one of the simplest methods, sufficient to perform well enough in many applications.
\par
Despite its simplicity, CCM is only reliable when its settings for light pulse shape integration are appropriately selected, and adjusting these settings can be time-consuming and error-prone.
Our aim was to develop tools and procedures for the standard Charge Comparison Method to enhance the scintillator characterization process, making it faster and more reliable. In the course of our work, we arrived at some interesting conclusions and identified opportunities to enhance PSD itself, leveraging the advantages offered by digital technology.

\section{Experimental details}

\subsection{Experimental setup}

\begin{figure}[H]
    \centering
    \includegraphics[width=0.8\linewidth]{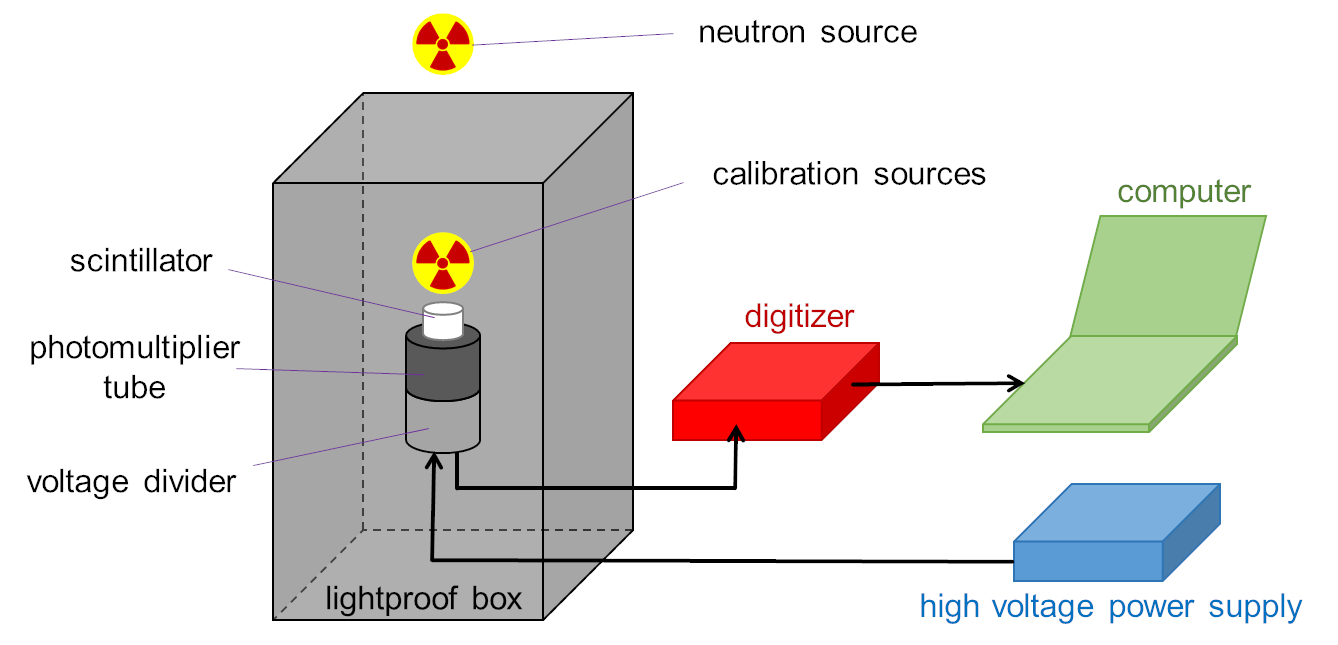}
    \caption{The layout of experimental setup used in neutron-gamma discrimination measurements.}
    \label{fig:experimental_setup}
\end{figure}

In the experiment, we used two scintillators: a \mbox{trans-stilbene} organic crystal and an organic glass scintillator (OGS).
Both were cylindrical, measuring 25.4~mm (1 inch) in height and diameter. 
The \mbox{trans-stilbene} is known for its excellent neutron-gamma discrimination capability, but exhibits a long scintillation decay time. In contrast, the organic glass shows poorer discrimination capability, but its decay time is significantly shorter. Both scintillators were wrapped in a white Teflon tape to maximize light collection and transfer to the photomultiplier tube (PMT). They were coupled with 2-inch diameter Hamamatsu \mbox{R6231-100} PMT \cite{R6231datasheet} using silicone grease (Baysilone Ol M of viscosity 1,000,000~mm\tsup{2}/s). The scintillator and photomultiplier were covered with black adhesive tape and placed in a lightproof box. A precisely regulated high-voltage power supply was used to power the PMT via a voltage divider. The anode output was fed to a CAEN DT5730 digital analyzer (digitizer) \cite{CEANwebsite}. We used the shortest possible cables between the photomultiplier and the digital analyzer to avoid signal distortions. The digitizer was controlled by CoMPASS software on a computer connected via USB cable (see Fig. \ref{fig:experimental_setup}).\par
\begin{figure}[H]
    \centering
    \includegraphics[width=0.9\linewidth]{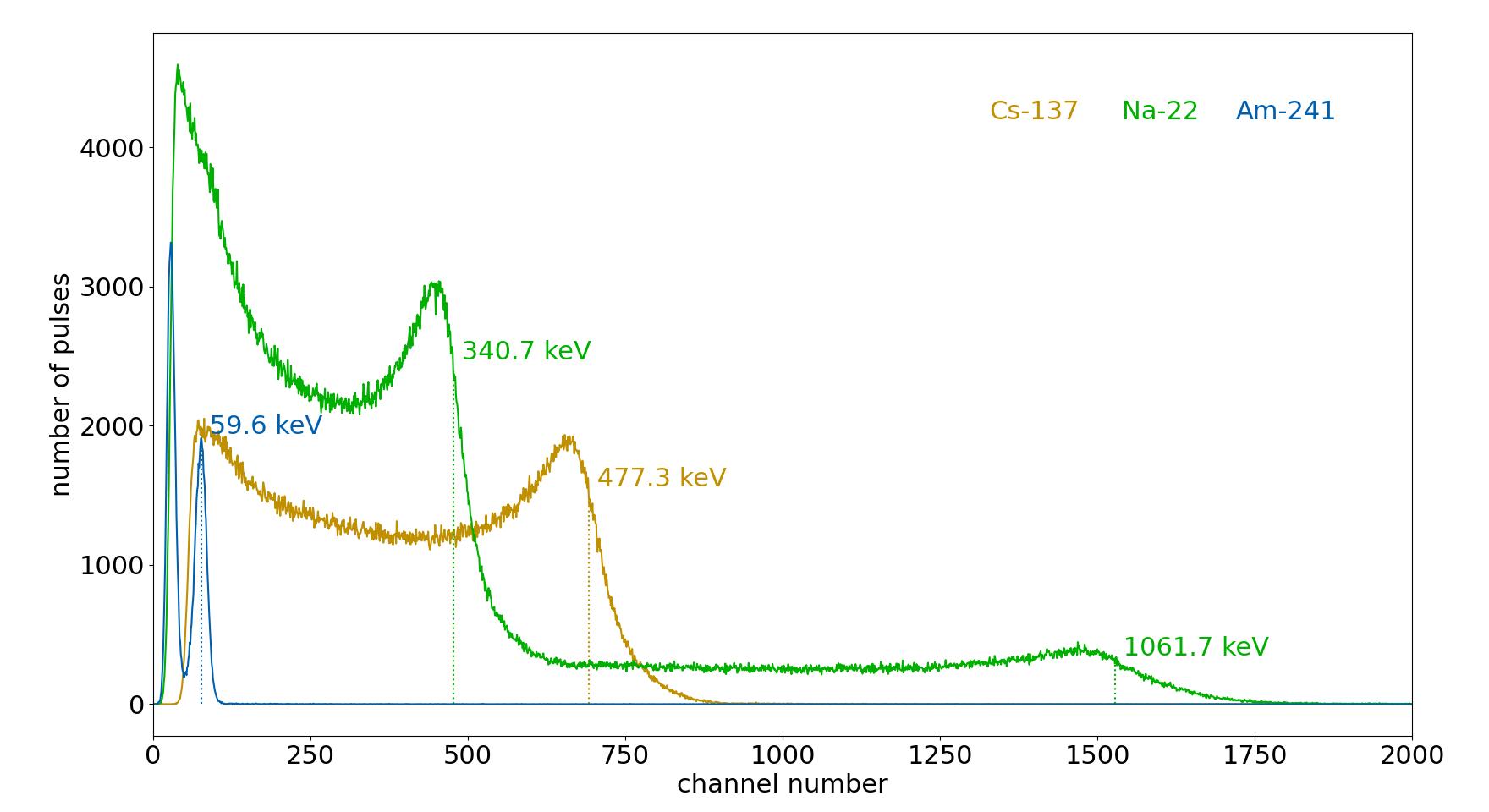}
    \caption{Energy spectra measured for OGS with three calibration sources: \tsup{137}Cs, \tsup{22}Na and \tsup{241}Am. Three Compton edges of 1061.7, 477.3 and 340.7~keV, and the 59.6~keV full-energy peak are depicted.}
    \label{fig:calibration_spectra}
\end{figure}

We used a PuBe neutron source to record the scintillator response to both neutron and gamma radiation of various energies. Additionally three different gamma sources were used for energy calibration: \tsup{241}Am (full energy peak 59.6~keV), \tsup{137}Cs (Compton edge 477.3~keV), \tsup{22}Na (Compton edges 340.7~keV and 1061.7~keV). Compton edges were estimated at the height 80\% of the maximum Compton distribution \cite{Swiderskietal2010}. Calibration sources were used inside the lightproof box right next to the detector due to their low radioactivity (less than 1000~kBq). Examples of calibration spectra measured for OGS are shown in Figure \ref{fig:calibration_spectra}. In measurements with mixed neutron/gamma radiation, the neutron source of activity of 24~GBq was placed outside the box. Polyethylene shielding around the neutron source was used to protect the personnel.

\subsection{Charge Comparison Method}

The scintillation pulses of an organic scintillator consist of several exponential components with different decay constants and intensities. The number of components and their duration depend mainly on the chemical composition of the scintillator, whereas the intensity is related to the nature of interactions with the measured radiation \cite{Knoll2010}. This means that different kinds of radiation may result in different time profiles of scintillation.\par
The idea of the Charge Comparison Method used in radiation discrimination is to quantify the total charge of the pulse and compare it with the charge corresponding to the tail part of the pulse. The tail encompasses the medium and slow components of the pulse. In some digital pulse processing devices, the tail is defined as the difference between the total charge ($Q_{long}$) and the initial part of the charge ($Q_{short}$) \cite{CoMPASSmanual}. This difference divided by the total charge makes the PSD parameter (see Eq. \ref{eq:PSD}).
The total charge is proportional to the total energy ($E$) deposited in the scintillator by the ionizing particle (Eq. \ref{eq:energy}).
In organic scintillators, only a small fraction of the particle energy is converted into light, and this fraction depends on the type of particle; e.g. 100-keV neutrons usually result in much lower energy deposited (via recoil protons) than 100-keV \mbox{gamma-ray} (via electrons). The experimental setup, depicted in Figure \ref{fig:experimental_setup}, did not allow a precise calibration of the neutron energy, and we used the keVee (keV electron equivalent) unit to express the light output produced by neutrons, where 1~keVee is by definition the light output by a 1-keV fast electron \cite{Knoll2010}. 

\begin{equation}
    \label{eq:PSD}
    PSD = \frac{Q_{long} - Q_{short}}{Q_{long}}
\end{equation}
\begin{equation}
    \label{eq:energy}
     Q_{long} \sim E
\end{equation}

\begin{figure}[H]
    \centering
    \includegraphics[width=0.8\linewidth]{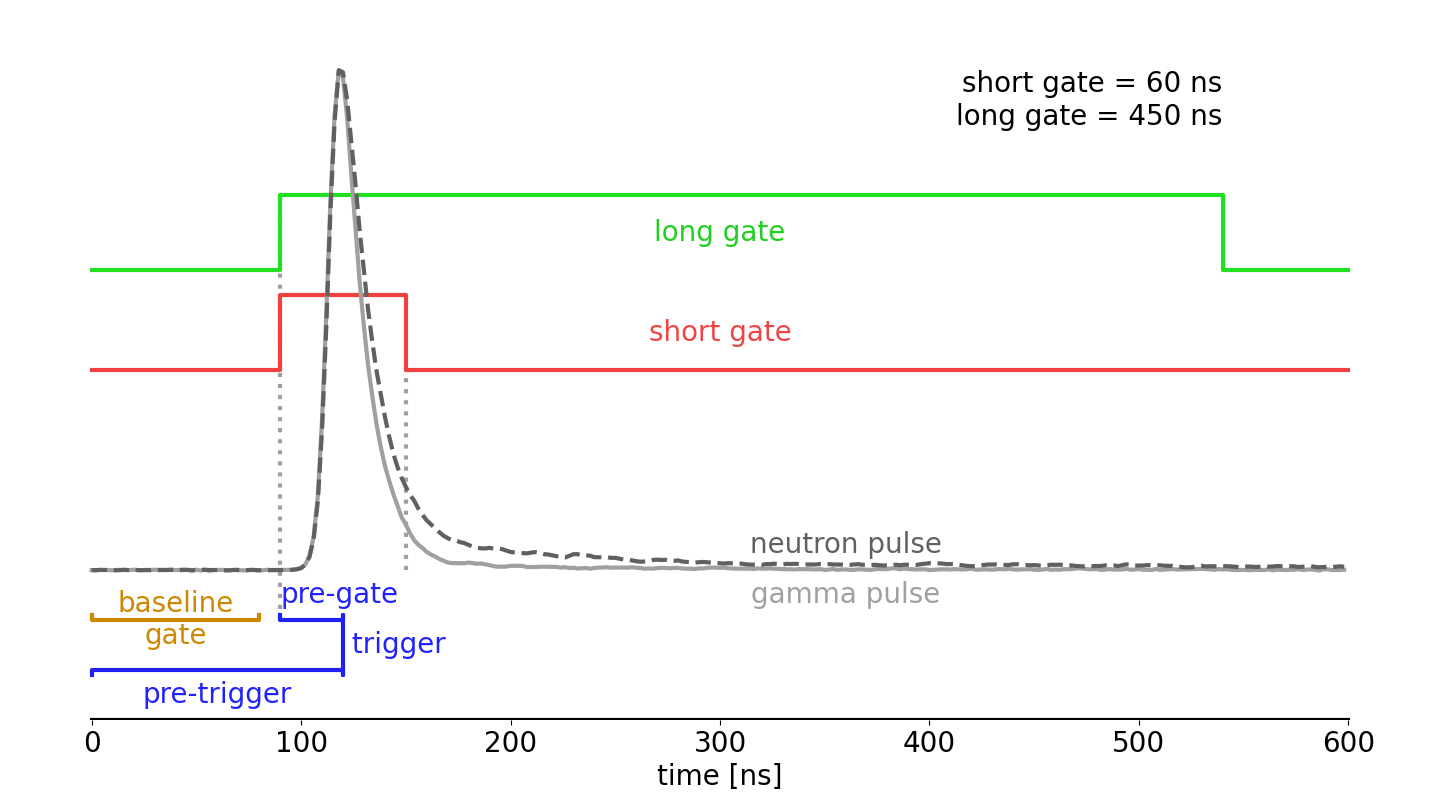}
    \caption{An example of gamma and neutron induced pulses in OGS recorded by a digital analyzer. The corresponding CCM parameters are: long gate, short gate and pre-gate. The pre-gate determines the start point of long- and short gate. The baseline is calculated as the average of pulse heights within the baseline gate.}
    \label{fig:pulses}
\end{figure}

In digital implementation $Q_{long}$ and $Q_{short}$ are integrals of a pulse in a long gate and a short gate, respectively (see Fig. \ref{fig:pulses}), where the lengths of these gates are usually preset manually.

\begin{figure}[H]
    \centering
    \includegraphics[width=0.7\linewidth]{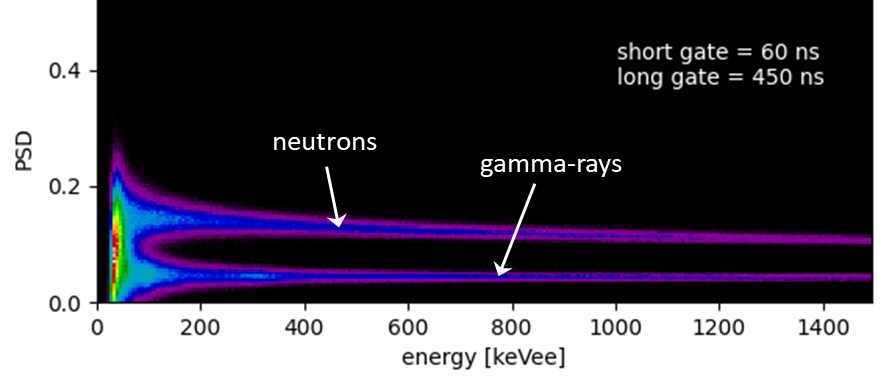}
    \caption{An example of 2D heatmap of the PSD parameter versus total charge (energy deposition) for OGS measured with a PuBe neutron source. Two branches of neutron and \mbox{gamma-ray} induced events are clearly separated.}
    \label{fig:PSDvsE}
\end{figure}

\begin{figure}[H]
    \centering
    \includegraphics[width=0.7\linewidth]{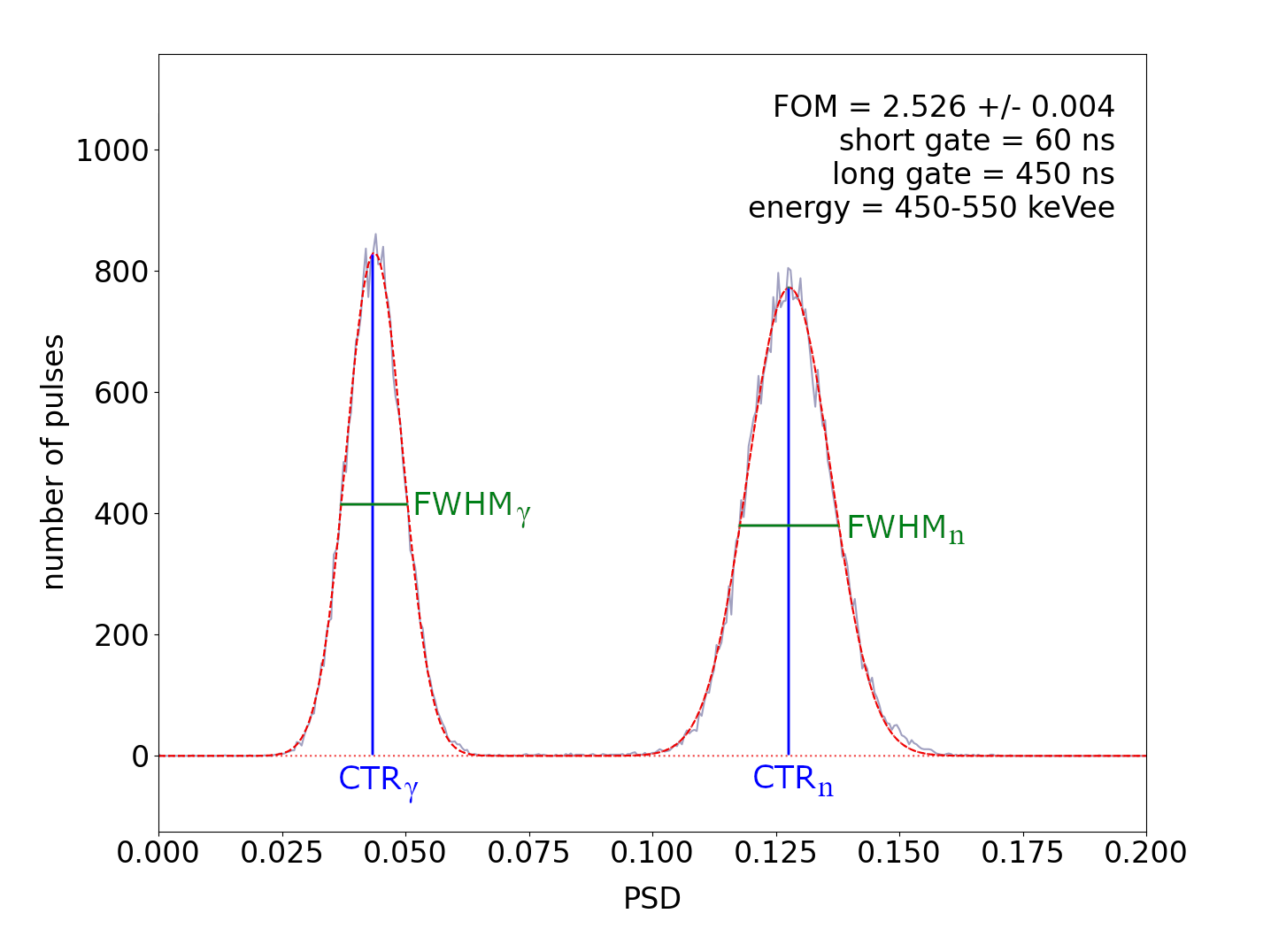}
    \caption{The projection of the 2D heatmap onto the PSD axis at 500$\pm50$~keVee energy cut for OGS measured with a PuBe neutron source. Solid gray line represents the raw data and red dashed line is the result of fitting two Gaussian peaks. Centroids (CTR) and Full Widths at Half Maximum (FWHM) of both peaks are depicted. Note: the Figure of Merit uncertainty here is strictly numerical, it does not include external factors. Real uncertainty is much higher.}
    \label{fig:PSDhist}
\end{figure}

Figure \ref{fig:PSDvsE} shows a 2D heatmap of the PSD parameter versus total charge (energy deposition) for the OGS sample. Two distinct groups of pulses are observed: one from neutrons and one from \mbox{gamma-rays}. Pulses from neutrons have longer tails and therefore, their PSD parameter has a higher value than in case of gamma rays. The projection of the 2D heatmap onto the PSD axis at 500$\pm50$ keVee energy cut for the OGS measured with a PuBe neutron source is plotted in Fig. \ref{fig:PSDhist}. 
Each of the peaks represents pulses from different kinds of radiation (neutrons and \mbox{gamma-rays}), and the performance of neutron-gamma discrimination can be quantified by calculating the Figure of Merit (FOM) defined by eq. \ref{eq:FOM}. 
\begin{equation}
    \label{eq:FOM}
    FOM = \frac{|CTR_{n}-CTR_{\gamma}|}{FWHM_{n}+FWHM_{\gamma}}
\end{equation}
FWHM and CTR stand for the Full Width at Half Maximum and the Centroid of the peaks, respectively, corresponding to neutron ($n$) and gamma~ray ($\gamma$) detection as projected onto the PSD parameter axis.
The FOM value is useful for comparing the discrimination capabilities of scintillators. However, it depends on the initial setting of the lengths of the long and short gates and on the \mbox{pre-gate} (the moment the gates start). Therefore, it is essential to choose these parameters thoughtfully to ensure an accurate comparison.

\subsection{Software} \label{sec:software}

We have developed a custom software in Python programming language to analyze raw pulses recorded with the CAEN DT5730 digital analyzer and the dedicated CoMPASS DAQ software. There are two additional parameters that must be taken into account when dealing with raw pulses: \mbox{pre-trigger} and the baseline.
\mbox{Pre-trigger} determines the time before the hardware trigger the data recording starts.
In Section \ref{sec:trigger}, we describe more details of the triggering methods and their impact on the final 3D and 2D spectra.
\par
Recorded waveforms are a raw output of the hardware analog-to-digital converter (ADC), and in calculating $Q_{long}$ and $Q_{short}$ integrals a baseline is used which refers to the DC voltage level named "zero level". The raw output signals have to be preprocessed according to their polarity and offset to obtain positive zero-based pulses that are CCM ready. The baseline is usually calculated dynamically by the CEAN digitizer for each pulse; however, it can also be set to a fixed value. We used the raw data from the waveform preceding the actual pulse to calculate the baseline (see the baseline gate in Fig. \ref{fig:pulses}).
\par
We used the NumPy library to manipulate the data and \textit{curve\_fit} function from SciPy library to estimate CTR and FWHM (this function implements the Levenberg-Marquardt algorithm \cite{SciPycurvefit}). This version of software had no proper user interface, but we used the matplotlib library to export the results in graphical form. The software was designed to be supervised and all crucial results to be revised by the user. The graphical representation of data turned out to be a good way to quickly find potential errors and their sources, as well as to see interesting patterns.
\par

\begin{figure}[H]
    \centering
    \includegraphics[width=0.98\linewidth]{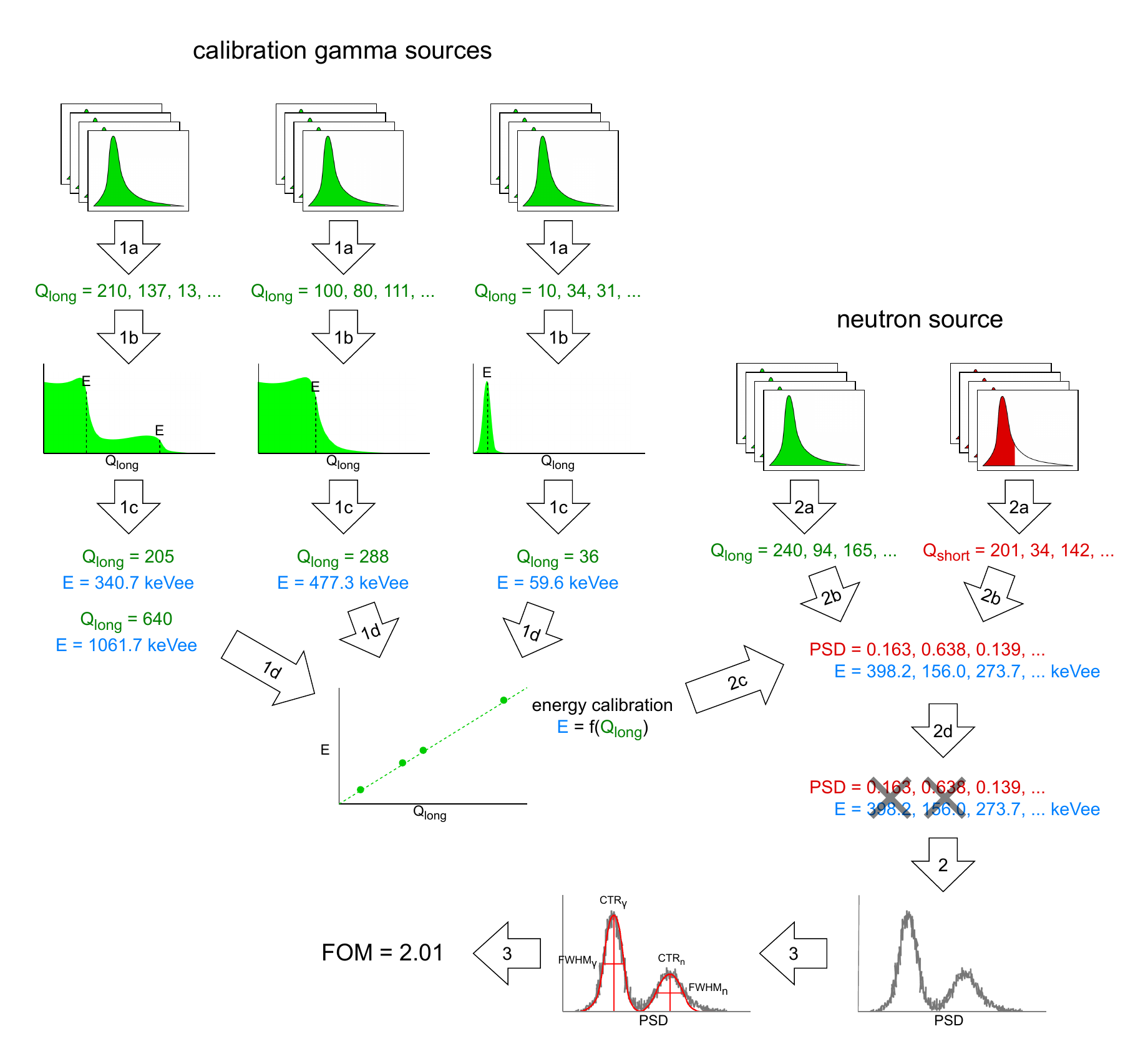}
    \caption{The idea of the algorithm used to process the raw data and calculate FOM automatically. Numbers and letters in the arrows refer to algorithm steps described in the text.}
    \label{fig:algorithm}
\end{figure}
The idea of the algorithm we used is simple (see Fig. \ref{fig:algorithm}). For a given set of parameters (long gate, short gate, \mbox{pre-gate}, \mbox{pre-trigger}, baseline gate):
\begin{enumerate}
    \item make energy calibration:
    \begin{enumerate}
    \item calculate $Q_{long}$ for each pulse (for each gamma-ray energy from a calibration source),
    \item make a histogram of $Q_{long}$ values (energy spectrum),
    \item assign the $Q_{long}$ value to the given energy $E$ from the calibration source,
    \item repeat previous steps for every calibration source to get the energy calibration,
    \end{enumerate}
    \item make PSD parameter histogram of pulses from a given energy range:
    \begin{enumerate}
    \item calculate $Q_{long}$ and $Q_{short}$ for each pulse from a neutron source,
    \item calculate PSD of each pulse from a neutron source using both $Q_{long}$ and $Q_{short}$ values,
    \item apply the energy calibration to determine $E$ of each pulse from a neutron source,
    \item select PSD of pulses of the energy $E$ in the given energy range to get the histogram,
    \end{enumerate}
    \item fit a Gaussian function to each peak in the PSD histogram and calculate the FOM value.
\end{enumerate}
By repeating this steps for a wide range of long and short gates, we could get a comprehensive amount of data to not only choose optimal gates with maximum FOM values, but also investigate some scintillator characteristics and pulse processing irregularities. 
\par
\begin{figure}[H]
    \centering
    \includegraphics[width=0.49\linewidth]{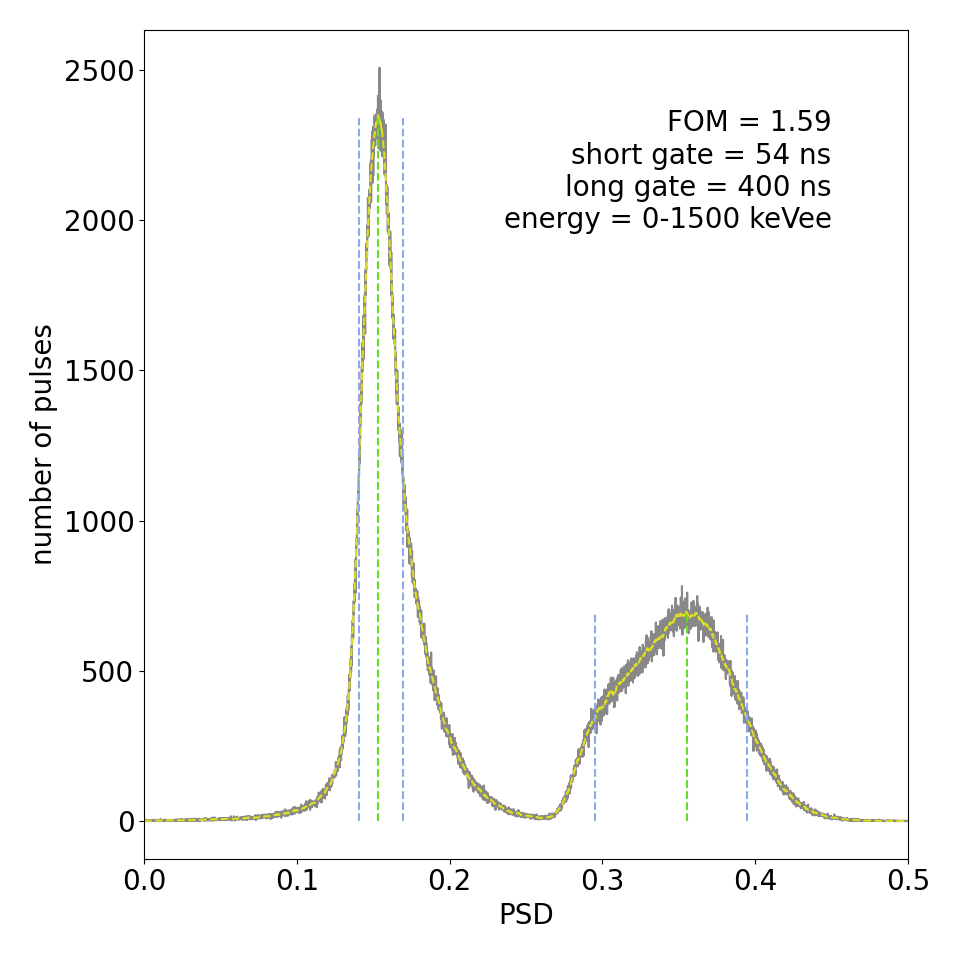}
    \includegraphics[width=0.49\linewidth]{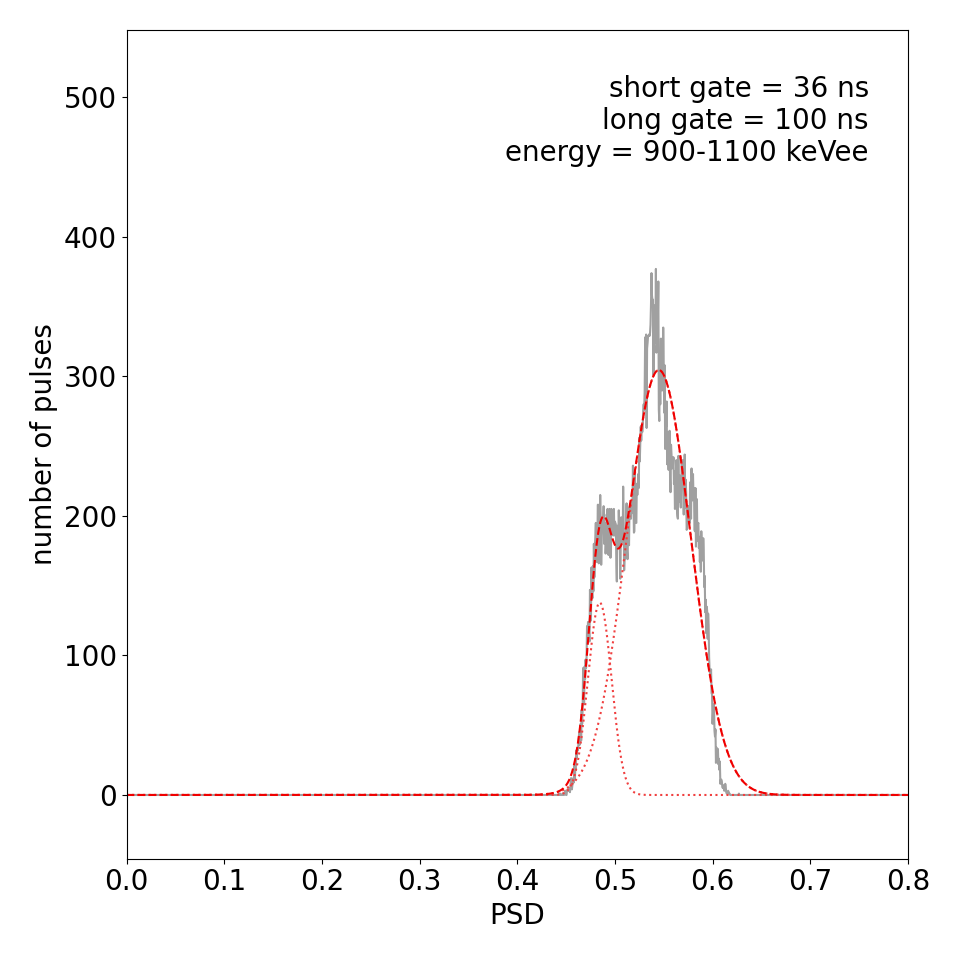}
    \caption{Examples of non-Gaussian PSD peaks in case of broad energy range (left) and extreme gates (right). Raw data are shown as gray plot. Yellow dashed line represents raw data smoothed with moving average. Vertical green lines are peaks maxima and vertical blue lines represent half height points of both peaks calculated by a method alternative to two Gaussian peaks fitting. Red dashed lines are the result of two Gaussian peaks fitting procedure, while red dotted lines are individual fitted peaks.}
    \label{fig:PSD_nongaussian}
\end{figure}

It is worth mentioning that the last step (3) of the algorithm is not a trivial task for a number of reasons:
\begin{itemize}
    \item statistical noises\\
    Depending on the total amount of data or how narrow the energy range is, the PSD histogram can have a significant amount of statistical noise that influences the process of fitting Gaussian peaks. The same applies to the energy spectra of calibration sources. Fortunately, this problem can be solved by making high-statistics measurements and choosing wide energy ranges with a large number of pulses.
    \item broad range of PSD values\\
    They can vary from below 0.01 to above 0.9 depending on the combination of short and long gates, as well as the scintillator itself. Usually, the more extreme these values, the more narrow the peaks, but there is no clear relation between the position of peaks on a histogram and their width. This presents another challenge because histogram intervals should be small enough to correctly represent the shape of both peaks and large enough not to make the aforementioned statistical noises problematic.
    \item peaks can overlap\\
    The more they overlap, the harder it is to fit a double Gaussian function to them. We use the \mbox{multi-step} process to overcome this challenge, approaching the correct fitting function parameters. This solution works quite reliably down to the FOM values around 0.5, however, below 0.8 the results need manual validating.
    \item peaks can be non-Gaussian\\
    This usually happens when a wide range of energies is taken into account or for extreme sizes and positions of gates. In the first case, we try to estimate the CTR of the peaks as their maximum position, and their FWHM is calculated as real width at a height of half that maximum (see Fig. \ref{fig:PSD_nongaussian} left). In the second case, when the PSD histogram is substantially distorted, it is impossible to tell neutron pulses from gamma pulses (Fig. \ref{fig:PSD_nongaussian} right). 
    \item flaws of iterative optimization algorithms\\
    Algorithms like the Levenberg-Marquardt one tend to find local maxima instead of global ones, so there is always possibility that the final fit is poor, and it is advised to check calculated values for obvious errors.
\end{itemize}

The software has been designed to be supervised and all crucial results to be revised by user. The graphical representation of data turned out to be a good way to quickly find potential errors and their sources, as well as to see interesting patterns. It is worth to notice that the overall time spent on the data acquisition is now significantly reduced (from days to hours), as a single run is enough to get more results than several measurements performed with parameters manually adjusted.
The time spent on data analysis did not remarkably increase.

\section{Results}

\subsection{Gates lengths} \label{sec:gates}

\begin{figure}[H]
    \centering
    \includegraphics[width=0.49\linewidth]{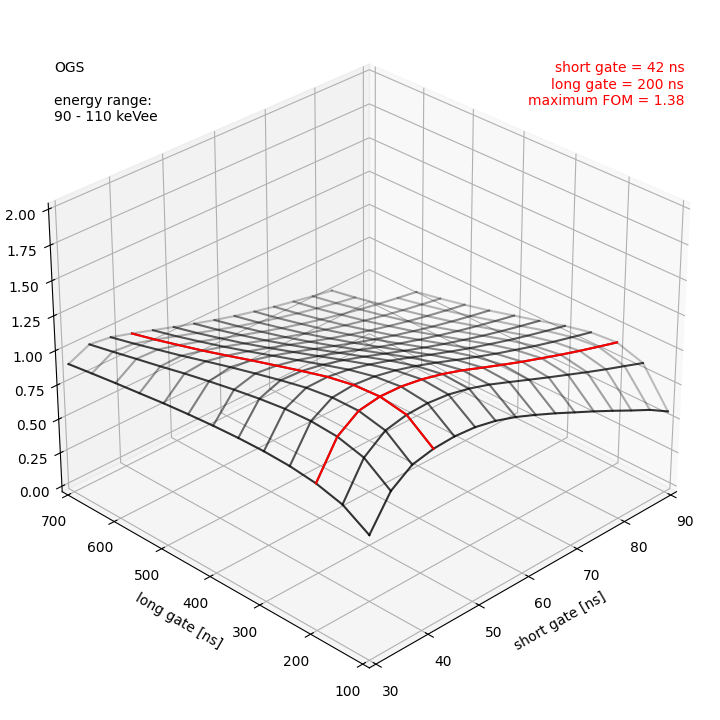}
    \includegraphics[width=0.49\linewidth]{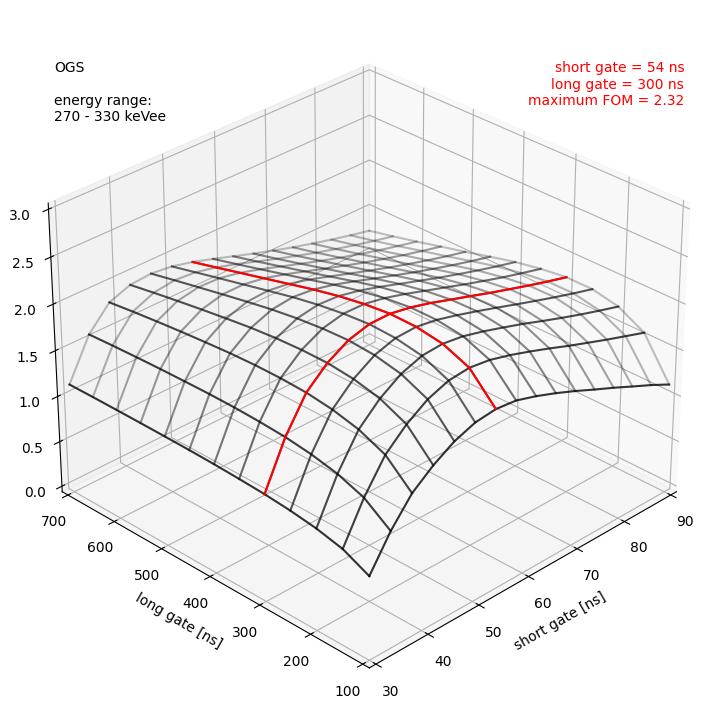}
    \includegraphics[width=0.49\linewidth]{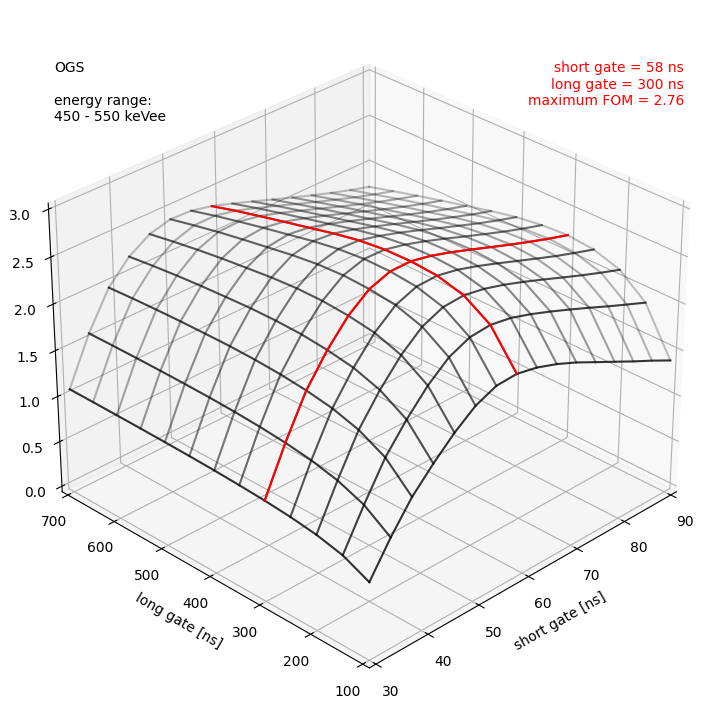}
    \includegraphics[width=0.49\linewidth]{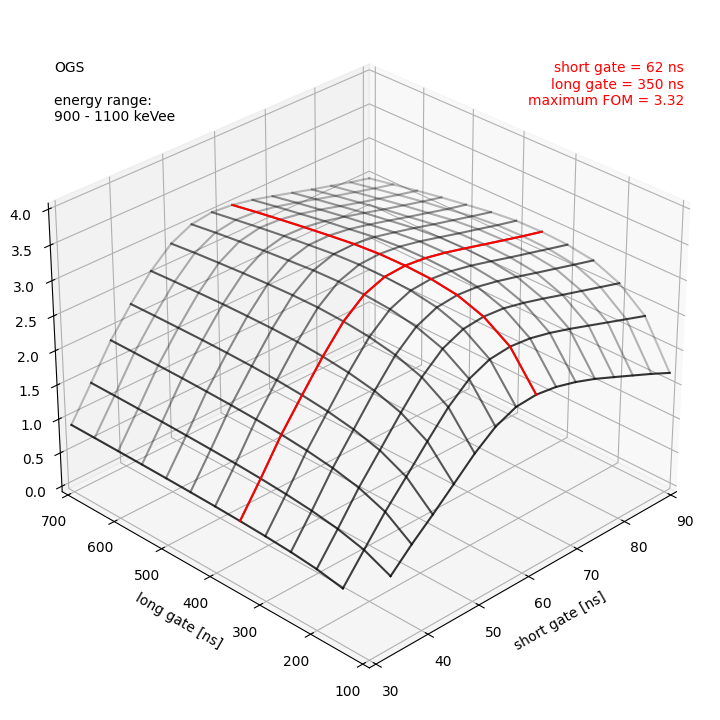}
    \caption{3D plot of FOM values for OGS. Gates with maximum FOM value for a given energy range are marked in red. The long gate was scanned with a step of 50~ns and short gate with 4~ns. Some points were manually removed because their FOM values were not reliable or false.}
    \label{fig:FOM3D_OGS}
\end{figure}

\begin{figure}[H]
    \centering
    \includegraphics[width=0.49\linewidth]{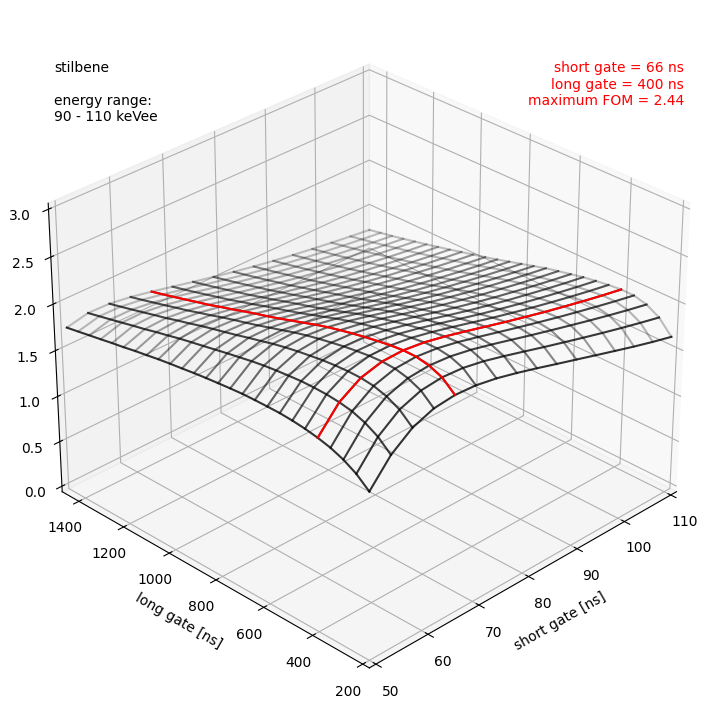}
    \includegraphics[width=0.49\linewidth]{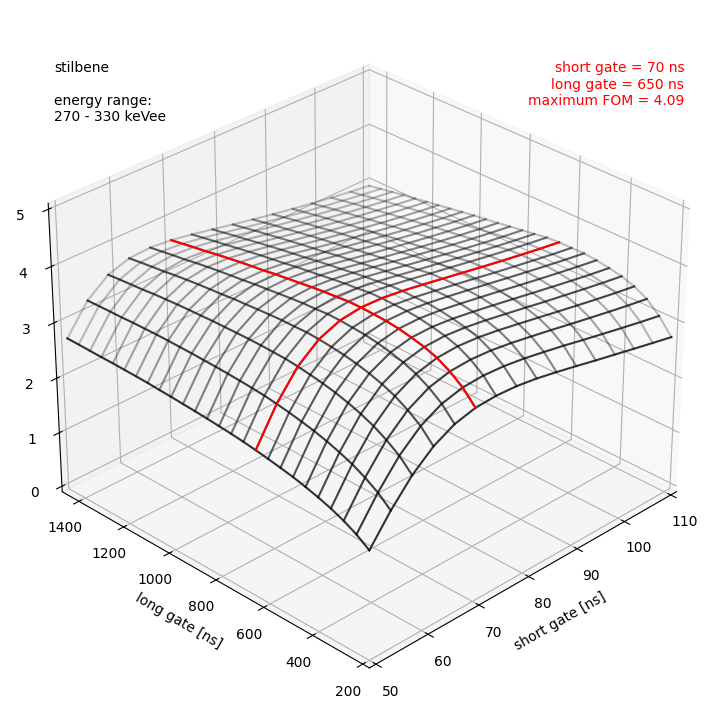}
    \includegraphics[width=0.49\linewidth]{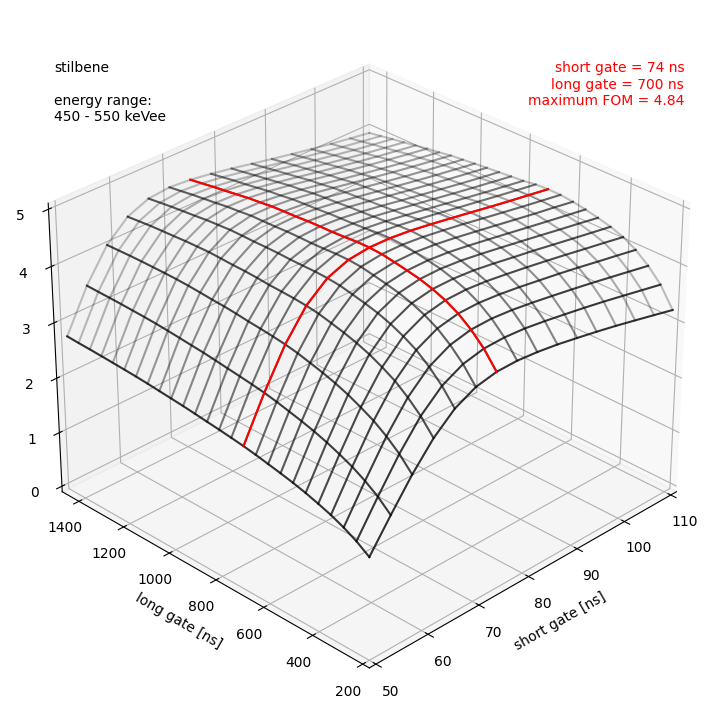}
    \includegraphics[width=0.49\linewidth]{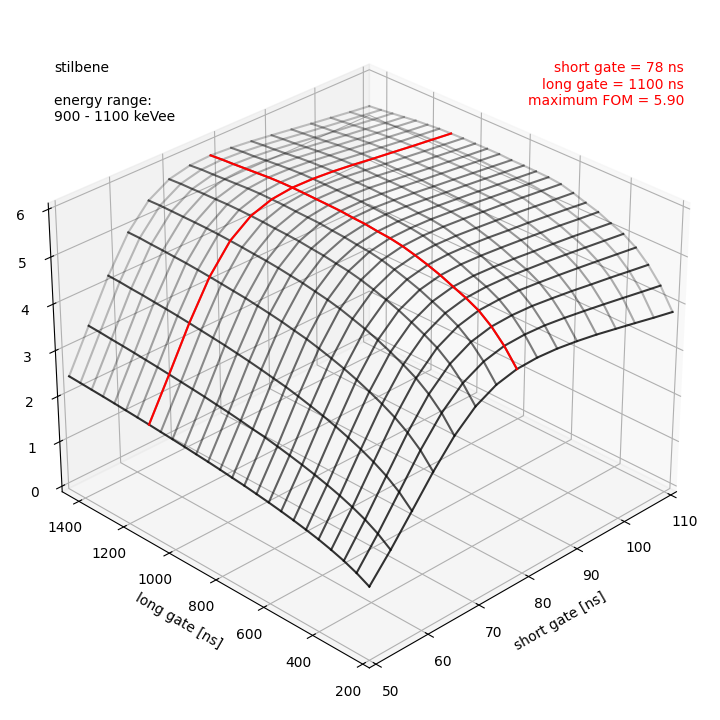}
    \caption{3D plot of FOM values for \mbox{trans-stilbene}. Gates with maximum FOM value for given energy range are marked in red. Long gate was scanned with a step of 50~ns and short gate with 4~ns.}
    \label{fig:FOM3D_stilbene}
\end{figure}

The lengths of the gates are the most important factors that determine the quality of CCM discrimination. When comparing two scintillators, the main focus is usually on finding these optimal gates and the best FOM values. First, the optimal values of the gates are estimated in a series of measurements. Then, FOM for many energies is calculated. However, it is much easier to find optimal gates when the entire landscape i.e. 3D plot of FOM versus the short and long gates can be seen instead of only a few calculated points. It is possible to implement a search tool in the software, but the goal was not only to find optimal values with the highest FOM, but also to be able to see the overall landscape, as shown in Figures \ref{fig:FOM3D_OGS} and \ref{fig:FOM3D_stilbene}. Therefore, we analyzed the data by systematically calculating the FOM for each pair from a broad range of short and long gates, similarly to brute-force searching.
\par
In many scientific papers, the FOM values were calculated in four specific energy ranges: 100\textpm10 keVee, 300\textpm30 keVee, 500\textpm50 keVee, and 1000\textpm100 keVee. We decided to use this $E \pm 0.1 \times E$ pattern instead of a fixed range width because the number of pulses decreases with the energy and choosing a fixed width would result in a lower number of pulses for higher energies and therefore higher statistical noises. Choosing $0.1 \times E$ keeps the results more reliable and comparable for different energies. We observe that for each energy range, there is a different pair of short and long gate lengths for which the Figure of Merit reaches its maximum (best) value. 

As seen in Figures \ref{fig:FOM3D_OGS} and \ref{fig:FOM3D_stilbene}, the FOM values decrease when the short gate length exceeds a certain optimal value for a given energy. This is because the short gate starts to embrace the tail part of the pulse which does not differ much for gamma and neutron pulses.
On the other hand, when the short gate is too short, it covers only the beginning of a pulse, which has a similar shape for \mbox{gamma-rays} and neutrons.
\par
Similar effect is also observed for a long gate, but the explanation is different. Due to electrical noise, the FOM degradation was found at the higher value of the long gate. The longer the gate, the more background noise is included in the integral of the pulse. Making the gate longer gave results similar to adding artificial Gaussian noise to the recorded waveforms. In both cases FWHMs of both PSD peaks are increasing (FWHM of gamma peak more, FWHM of neutron peak less), and therefore FOM is decreasing. In the lower-energy region the amplitude of pulses is closer to the amplitude of the noise. Therefore, maximum FOM values also decrease with energy and are superior at narrower long gates. It was also checked that smoothing the acquired raw waveforms with a simple moving average did not improve \mbox{neutron-gamma} discrimination performance.

\subsection{Triggering method} \label{sec:trigger}

We measured each scintillator with two different triggering methods: Leading Edge (LE) and Constant Fraction Discrimination (CFD). We also applied a new experimental triggering method in the offline analysis of digitally recorded data, which is described in Section \ref{sec:trigger_maximum}.

\subsubsection{Leading Edge} \label{sec:trigger_LE}

Leading Edge is the simplest triggering method based on a fixed threshold, which has to be exceeded by a pulse to record the event. It has two major disadvantages as compared to other methods:
\begin{itemize}
    \item the threshold has to be high enough not to trigger pulse acquisition with random noises and low enough to measure required signal - this limits the energy range of measurement,
    \item the lower the pulse, the later the measurement is triggered (this is called ``amplitude walk'', see Fig. \ref{fig:triggers}) - this causes gate-dependent distortions of PSD vs. energy histogram (see Fig. \ref{fig:LEvsCFDvsMAX_OGS} and \ref{fig:LEvsCFDvsMAX_stilbene}.)
\end{itemize}
Leading Edge is known to have a small time jitter for a fixed pulse height, but for different heights the amplitude walk is dominant to the time jitter \cite{Knoll2010}. Moreover, in applications such as \mbox{neutron-gamma} discrimination, the time jitter is negligible.

\subsubsection{Constant Fraction Discrimination} \label{sec:trigger_CFD}

\begin{figure}[H]
    \centering
    \includegraphics[width=0.75\linewidth]{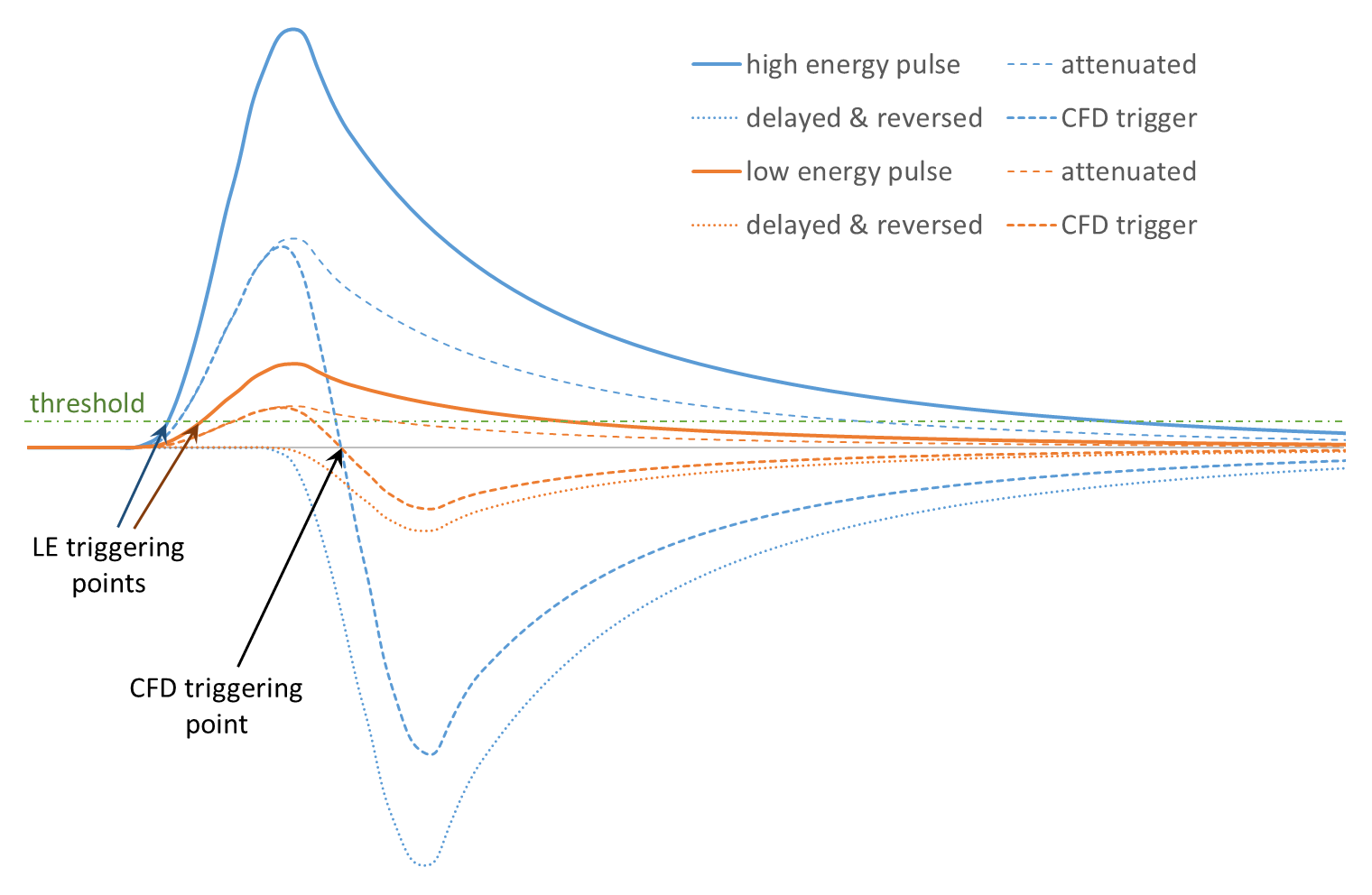}
    \caption{Comparison of Leading Edge (LE) and Constant Fraction Discrimination (CFD) triggering methods in case of high- and low-energy pulses. The low energy pulse triggering point of LE is delayed compared to the high energy pulse. The triggering point of CFD is the same regardless the height of the pulse.}
    \label{fig:triggers}
\end{figure}

Constant Fraction Discrimination is a more sophisticated method that performs a series of signal manipulations to estimate more precisely when a pulse reaches a certain fraction of its height. These manipulations include delaying the pulse, reversing, and attenuating it and summing attenuated and delayed pulses to find the zero-crossing time, which is independent of pulse amplitude (Fig. \ref{fig:triggers}) \cite{Knoll2010}, \cite{CoMPASSmanual}. This method can be implemented in an analog or digital way. However, in practice in both cases trigger is delayed compared to Leading Edge method due to the fact that signal manipulation takes a certain amount of time.
\par
It is also important to mention that there are several parameters in the CFD method (compared to one in LE) that must be tuned to obtain a reliable trigger. These parameters include the delay time and the attenuation factor. The delay time must be short enough not to produce multiple zero crossing points and long enough to produce a triggering signal significantly higher than electrical noise.
The attenuation factor is also limited by noise, because the triggering signal cannot be lower than a threshold, which is another parameter of the CFD implementation in CAEN digitizers \cite{CoMPASSmanual}.
\par

\subsubsection{Pulse Maximum} \label{sec:trigger_maximum}

The CFD technique aims to determine the exact time at which the pulse reaches a certain fraction of its height. In CAEN digitizers, the user can choose the fraction of 25\%, 50\%, 75\%, or 100\%.\footnote{Depending on settings and input signal length, the CAEN implementation of CFD \cite{CoMPASSmanual} works as one of the methods described by Knoll: Constant Fraction Timing or Amplitude and Rise Time Compensated Timing \cite{Knoll2010}. The meaning of the attenuation factor in each of these methods is slightly different and does not always represent the exact fraction of the pulse height.}
In case of the 100\% fraction, there is no need to apply all of these complicated data manipulations -- it is just the time the pulse reaches its maximum. With digitally recorded pulse shapes, it is as simple as finding the index of the maximum value in an array. A more sophisticated method of finding its maximum may be needed when the pulse is not smooth or has spikes.
\par
\begin{figure}[H]
    \centering
    \includegraphics[width=0.49\linewidth]{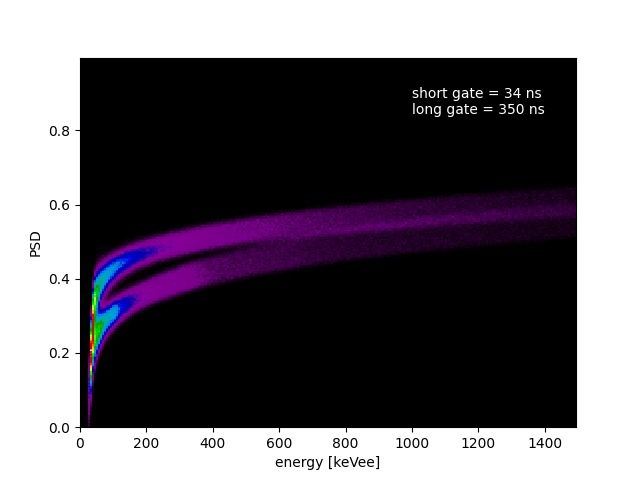}
    \includegraphics[width=0.49\linewidth]{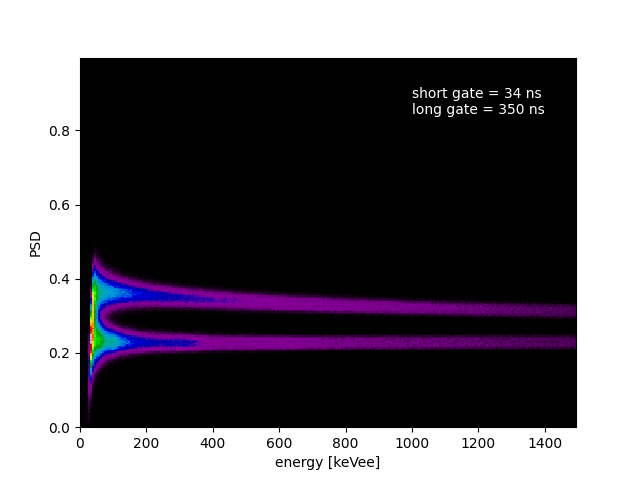}
    \includegraphics[width=0.49\linewidth]{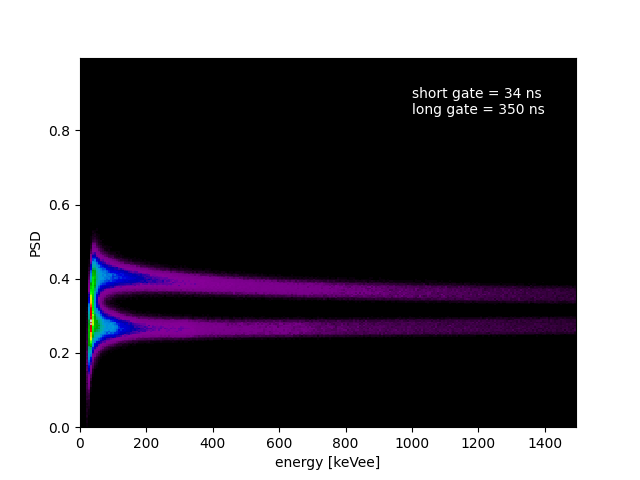}
    \includegraphics[width=0.49\linewidth]{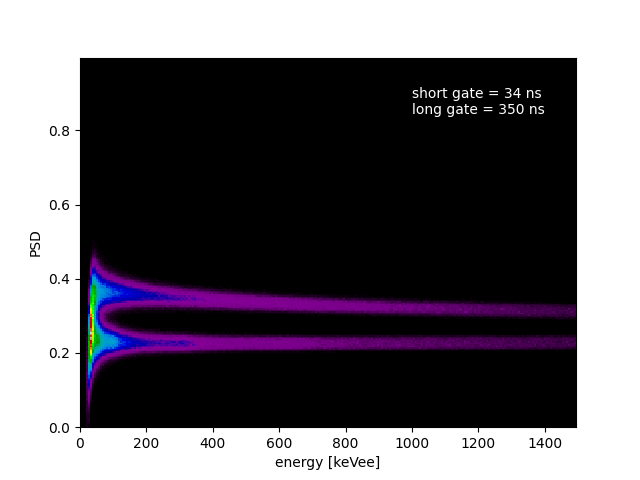}
    \caption{Example of OGS measurement data recorded with Leading Edge (top row) and Constant Fraction Discrimination (bottom row) triggering method. Data were analyzed with gates starting relative to the hardware trigger (left column) or the pulse maximum (right column). Note that suboptimal gates values are intentionally chosen to show a distortion of neutron and gamma branches, especially for the LE triggering mode (top left).}
    \label{fig:LEvsCFDvsMAX_OGS}
\end{figure}
\begin{figure}[H]
    \centering
    \includegraphics[width=0.49\linewidth]{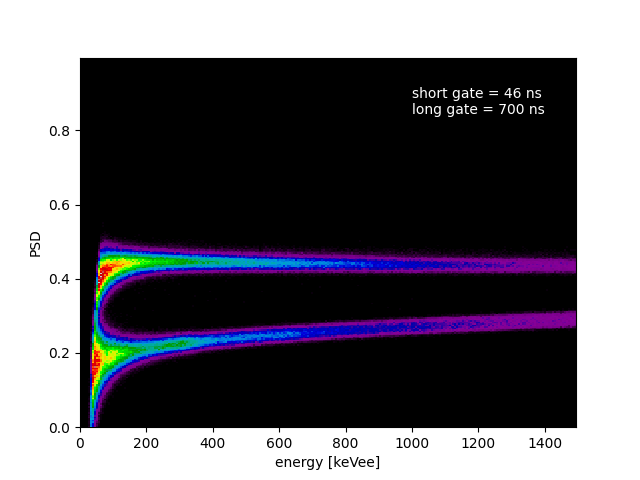}
    \includegraphics[width=0.49\linewidth]{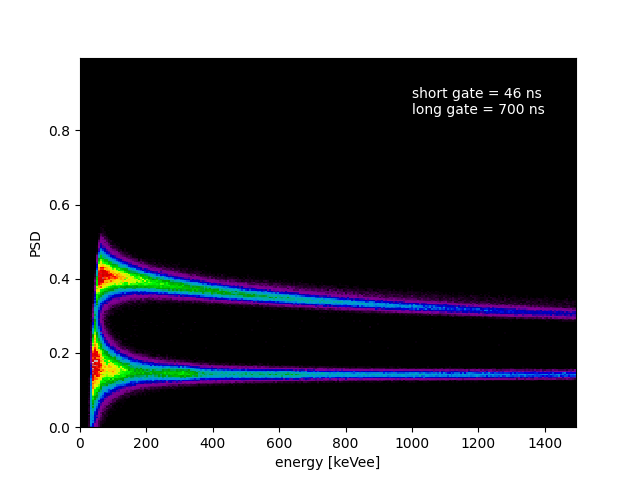}
    \includegraphics[width=0.49\linewidth]{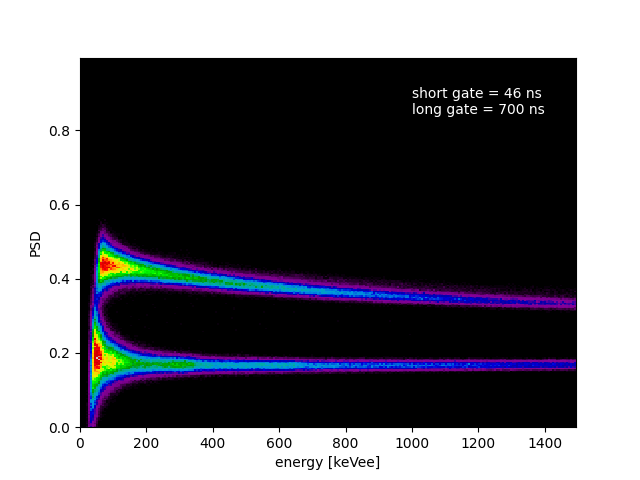}
    \includegraphics[width=0.49\linewidth]{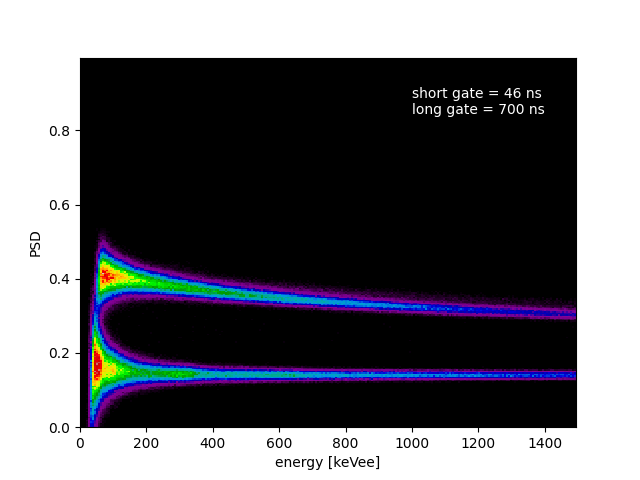}
    \caption{Example of \mbox{trans-stilbene} measurement data recorded with Leading Edge (top row) and Constant Fraction Discrimination (bottom row) triggering method. Data were analyzed with gates starting relative to the hardware trigger (left column) or the pulse maximum (right column). Note that suboptimal gates values are chosen to bring out problems that may arise due to triggering.}
    \label{fig:LEvsCFDvsMAX_stilbene}
\end{figure}

We decided to use the pulse maximum triggering in our software and apply it to the pulses recorded with the LE and CFD hardware triggering methods. The sample results are shown in Figures \ref{fig:LEvsCFDvsMAX_OGS} and \ref{fig:LEvsCFDvsMAX_stilbene}.
As expected, there is little or no change in the results between the CFD triggering (bottom left of Figures \ref{fig:LEvsCFDvsMAX_OGS} and \ref{fig:LEvsCFDvsMAX_stilbene}) and the pulse maximum triggering (bottom right). There is a small shift along the PSD axis because the gates start at different times. The importance of this starting point is discussed in Section \ref{sec:starting_point}.
Moreover, when the maximum of the pulse is applied as a trigger, the results from the data measured with LE triggering (top left of Figures \ref{fig:LEvsCFDvsMAX_OGS} and \ref{fig:LEvsCFDvsMAX_stilbene}) transform to almost the same 2D heatmap as the CFD data (top right).
\par
It is worth mentioning that in Figures \ref{fig:LEvsCFDvsMAX_OGS} and \ref{fig:LEvsCFDvsMAX_stilbene} we have deliberately chosen the \mbox{non-optimal} values of gates to bring out some problems that may arise when parameters are far from optimal. One of the problems is the overlap of neutron and gamma branches of the 2D heatmap that can be seen in the top left of Figure \ref{fig:LEvsCFDvsMAX_OGS}, i.e. for OGS data with LE triggering. The overlap can occur for both LE and CFD, but is more severe in the case of LE. This is one of the reasons why the triggering method, as well as the gates, should be chosen carefully when it comes to PSD discrimination.

\subsection{Gates starting point} \label{sec:starting_point}

As we have seen in Section \ref{sec:gates}, the lengths of the gates values are not universal, and therefore they cannot be considered scintillator properties. They are dependent on the time point at which both gates start. This point should be set before the pulse starts (at least if we want to include whole the pulse in the analysis), and much earlier that the triggering point. In CEAN digitizers, there is a parameter called ``pre-gate'' that determines the time between the gates starting point and the trigger (see Fig. \ref{fig:pulses}). Typically, it is manually set to a safe value to guarantee that the gates start at the right time. This is usually estimated first by viewing raw pulses. Digitally recorded pulses allow a user to test a range of different pre-gate values to see how it influences the neutron-gamma discrimination capability.
\par
\begin{figure}[H]
    \centering
    \includegraphics[width=0.49\linewidth]{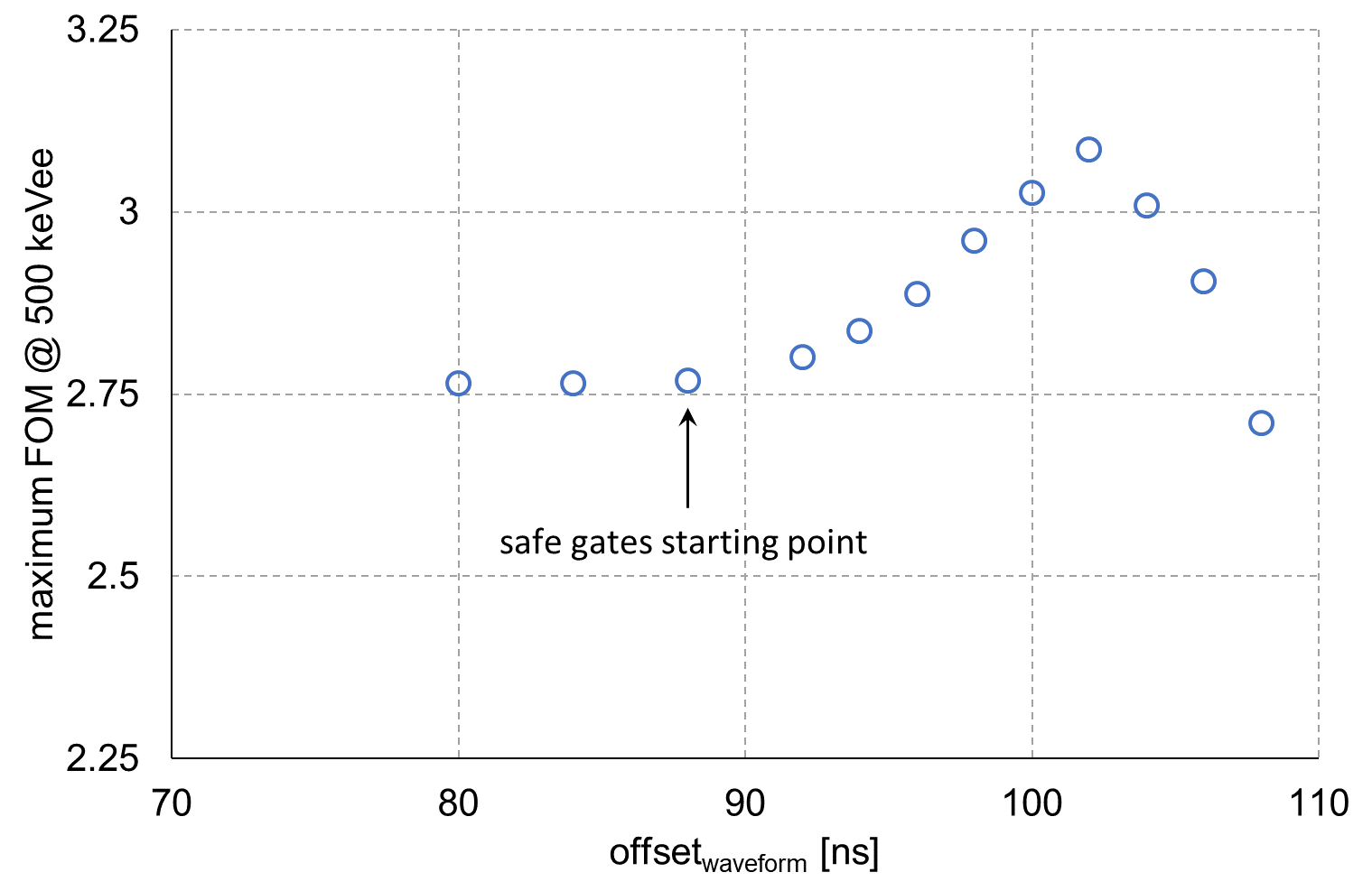}
    \includegraphics[width=0.49\linewidth]{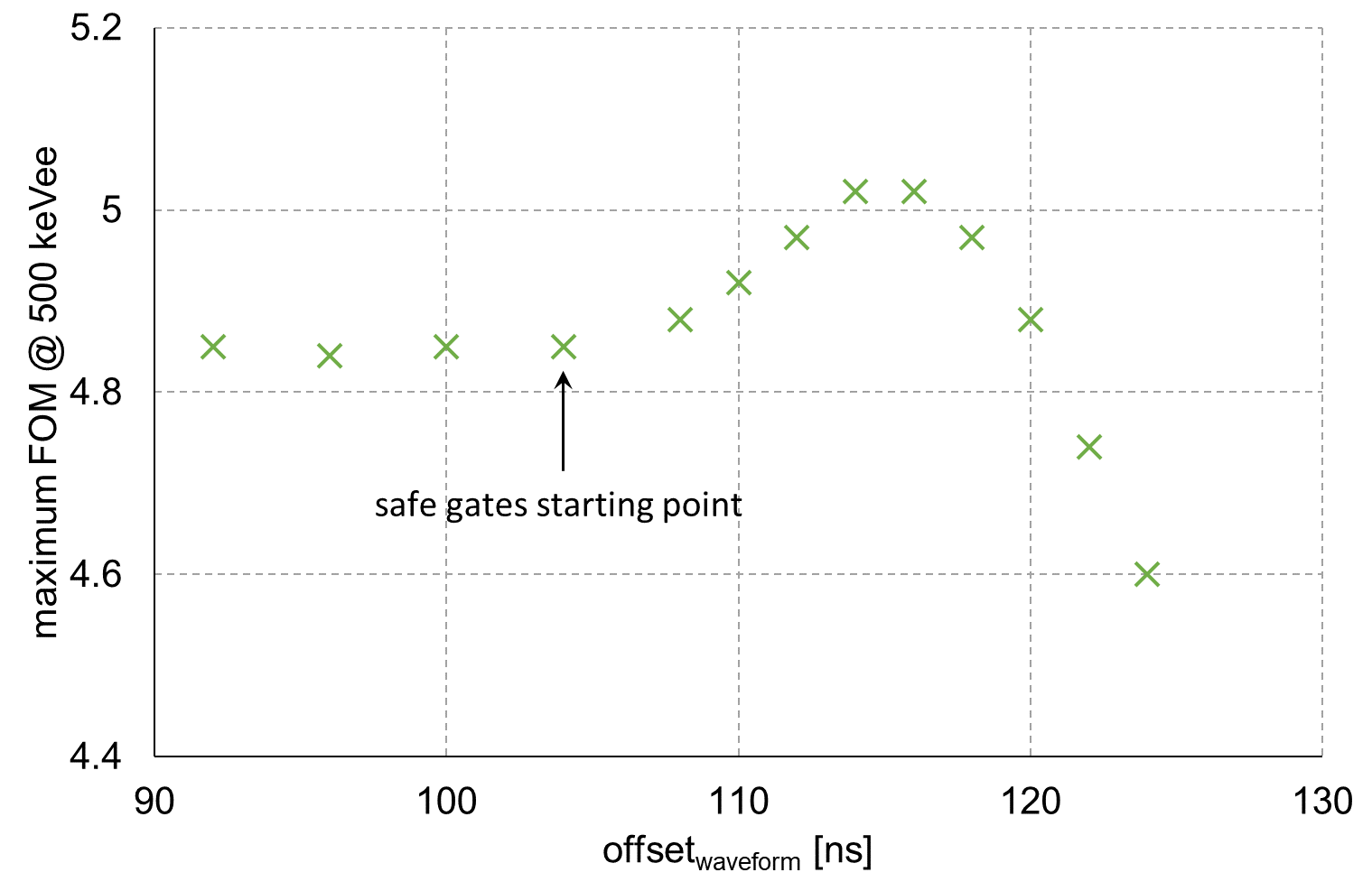}
    \caption{Changes of maximum value of FOM depending on \textit{offset\tsub{waveform}} (the offset between the starting point of gates and the beginning of the waveform) in case of OGS (left) and \mbox{trans-stilbene} (right) at energy 500~keVee. Starting point value depends on hardware settings (see values on horizontal axis). All FOM values were calculated for the same set of data for each scintillator, so their uncertainties are strictly numerical and too small to show in the graph.}
    \label{fig:starting_point_zero}
\end{figure}

In practice, it is more convenient to define the offset of the starting point relative to the beginning of the waveform (\textit{offset\tsub{waveform}}) rather than relative to the hardware trigger. In terms of CAEN digitizer settings \textit{offset\tsub{waveform} = \mbox{pre-trigger} -- \mbox{pre-gate}}. Figure \ref{fig:starting_point_zero} shows the dependence of maximum FOM values on this offset for both OGS and \mbox{trans-stilbene}. We can see that the FOM values do not change up to a certain value of \textit{offset\tsub{waveform}}. In examples shown in Fig. \ref{fig:starting_point_zero} these values are around 88~ns for OGS and around 104~ns for \mbox{trans-stilbene}. They are most likely the time values that the pulses start and it would be optimal to choose them as the safe gates starting point for further data analysis. However, these values are dependent on hardware settings and should be estimated for each measurement separately. Therefore, it would be even more convenient to find another way to estimate optimal gates starting point.
\par
\begin{figure}[H]
    \centering
    \includegraphics[width=0.49\linewidth]{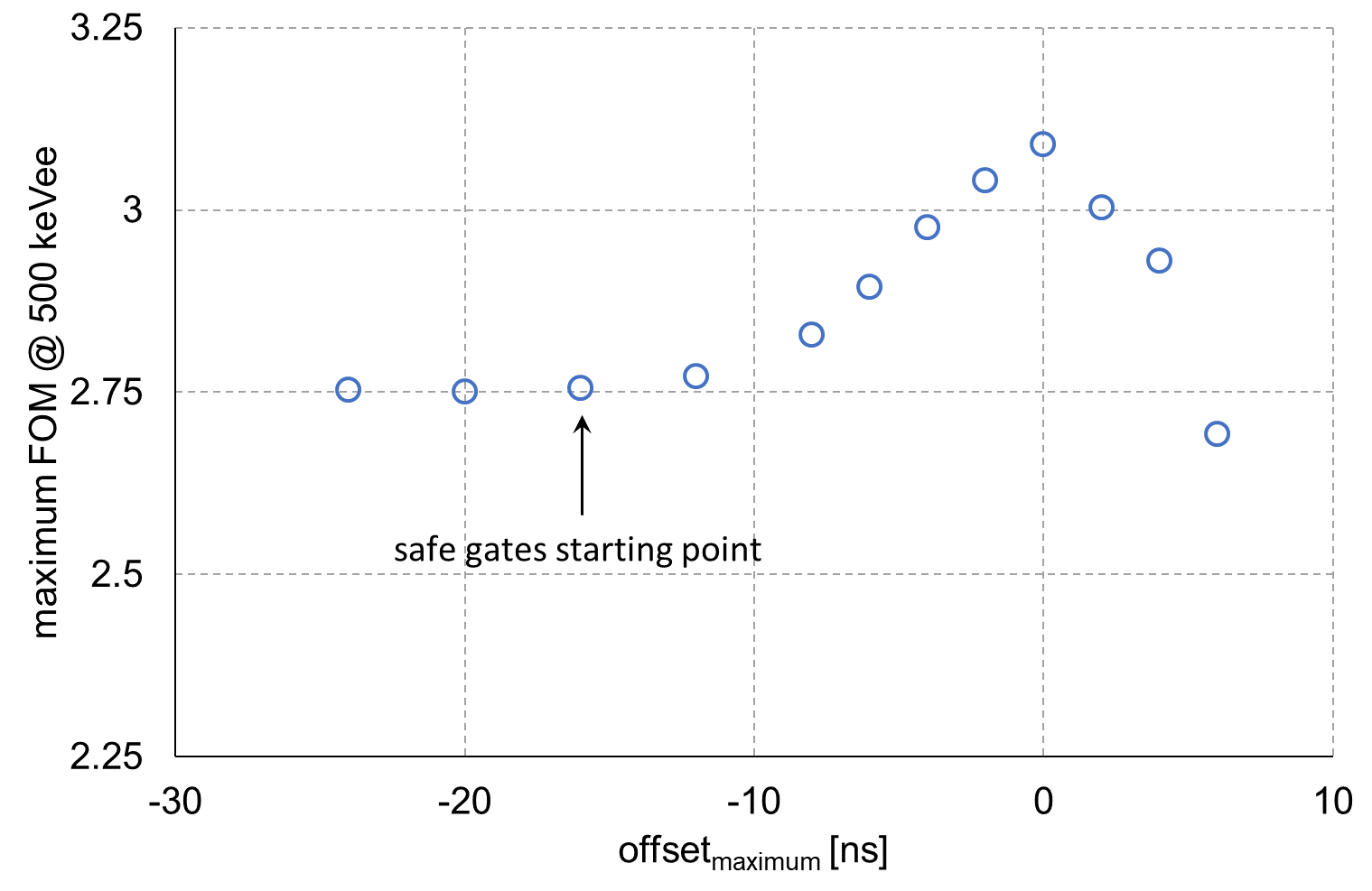}
    \includegraphics[width=0.49\linewidth]{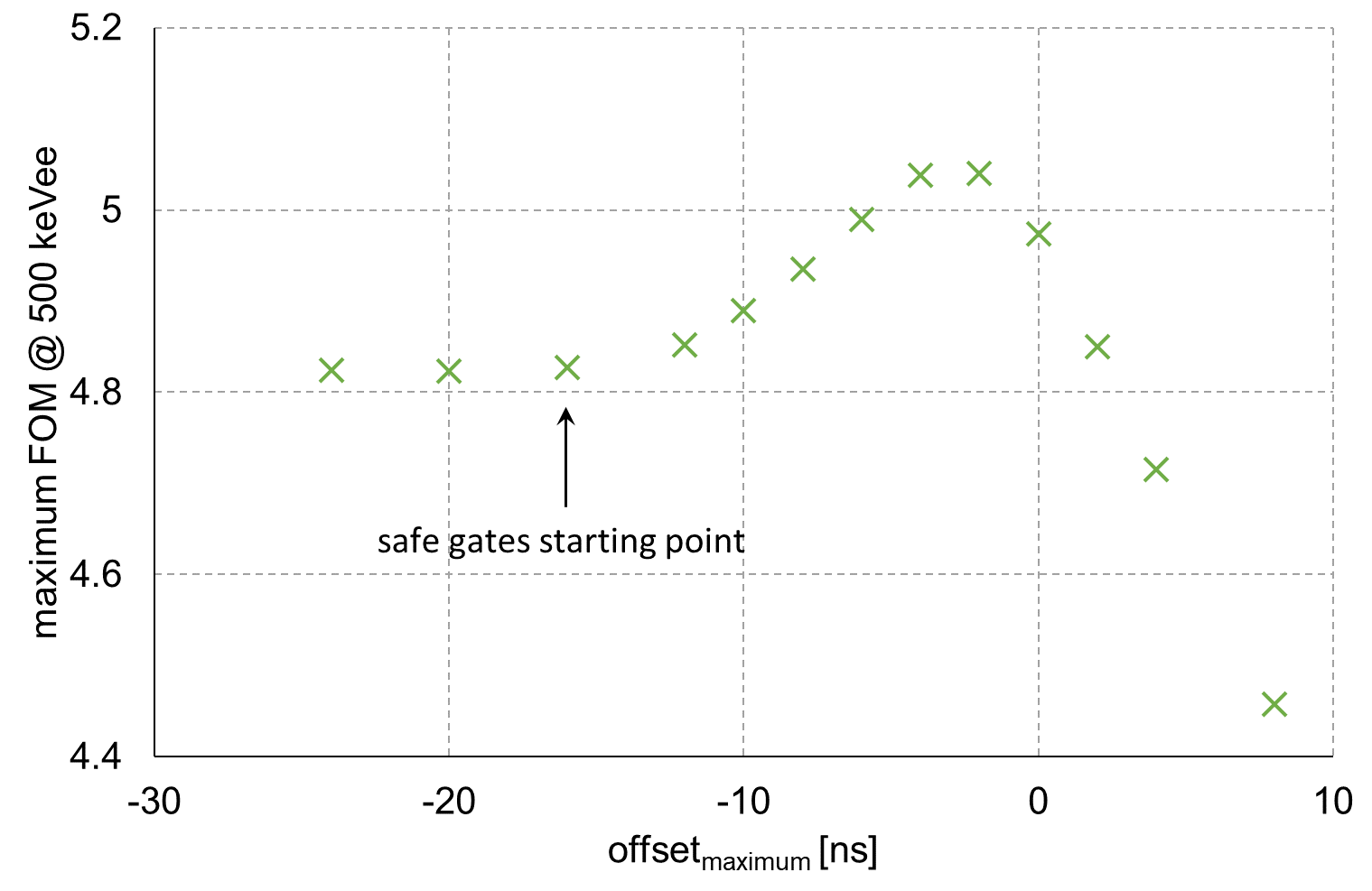}
    \caption{Changes of maximum value of FOM depending on \textit{offset\tsub{maximum}} (the offset between starting point of gates and the maximum of the pulse) in case of OGS (left) and \mbox{trans-stilbene} (right) at energy 500~keVee. Starting point value is independent on hardware settings (see values on horizontal axis). All FOM values were calculated for the same set of data for each scintillator, so their uncertainties are strictly numerical and too small to show in the graph.}
    \label{fig:starting_point_max}
\end{figure}

We analyzed the same data with another offset, defined relative to the pulse maximum (\textit{offset\tsub{maximum}}). Unlike \textit{offset\tsub{waveform}}, this one can have negative values (which refer to gates starting before the pulse maximum) and positive values (which refer to gates starting after the pulse maximum).
The results are in Fig. \ref{fig:starting_point_max}. There is little or no difference in the FOM values, and the curve is almost identical to that in Fig. \ref{fig:starting_point_zero}. What is different is that with the pulse maximum as a reference, it is much easier to estimate a universal and safe starting point value. In the cases of both measured scintillators, it is around 16~ns before the pulse maximum (\textit{offset\tsub{maximum}} = -16~ns) and is not dependent on any hardware settings. Moreover, the values of \textit{\mbox{pre-trigger}} and \textit{\mbox{pre-gate}} (needed to use \textit{offset\tsub{waveform}} effectively) can sometimes be lost, as both are not recorded automatically along with pulses and have to be saved independently, while the pulse maximum is an information already present in each waveform being analyzed.

Regardless of the method of setting the starting point, it is interesting to note that the FOM values are increasing when the starting point is in the range of the rising slope of the pulse and the FOM reaches a maximum value at a starting point around the maximum of the pulse. Then, they decrease as the starting point reaches the decreasing part of the pulse. This behavior is the same for the two scintillators that we used. However, it is more a data processing feature than a scintillator property.
\par
This is easy to understand when one notices that the rising slope is very similar in both neutron and gamma pulses (see Fig. \ref{fig:pulses}). Removing this part from the short gate subtracts the same value from $Q_{short}$ for both of them, making the differences between neutrons and \mbox{gamma-rays} in $Q_{short}$ relatively higher. As our results suggest, this increase in FOM can reach more than 10\% for OGS, but less than 5\% for \mbox{trans-stilbene} (Fig. \ref{fig:starting_point_zero}, \ref{fig:starting_point_max}).
\par
It is worth mentioning that the rising time of the pulse is dependent not only on the scintillator and radiation type, but also on the bandwidth of the hardware used to analyze the signal. In case of the CAEN digitizer we used, this bandwidth is limited to 250 MHz \cite{CEANwebsite}, so in digitizers with higher frequency limit this significant increase in FOM may not be present. On the other hand, this also means that in the hardware with much lower bandwidth, it may be beneficial to implement starting gates at pulse maximum. As mentioned in Section \ref{sec:trigger_maximum}, it is also a viable option as a trigger, so it could be of dual purpose without additional computational cost.

\subsection{Baseline} \label{sec:baseline}

The baseline estimation appeared to be another factor influencing PSD discrimination quality. We experienced some differences in $Q_{short}$ and $Q_{long}$ between our software and CAEN digitizer firmware. These differences resulted in lower FOM values obtained with our software, even if we tried to replicate the procedure that digitizer performs. There are several factors to be aware of, like number rounding or what part of signal is used to calculate the baseline. The firmware implementation of baseline calculation (unknown to us) seems to be different than method of off-line baseline calculation in digitally recorded pulses recommended by digitizer producer \cite{CAENsupport}. In fact, some of data used by digitizer to calculate the baseline may not be present in recorded pulses. Although the digitizer we used was able to output raw baseline instead of pulse shape, unfortunately, not both at the same time, so we couldn't easily diagnose what was the cause of these differences.
\par
On the other hand, baseline calculation in our software was consistent in all the data analysis we did, so our overall conclusions seem not to be affected significantly. The problem may occur when comparing scintillators characterized with different experimental setups and procedures, as significant differences in the results may not represent real differences between the scintillators accurately.

\subsection{FOM versus separation} \label{sec:separation}

\begin{figure}[H]
    \centering
    \includegraphics[width=0.45\linewidth]{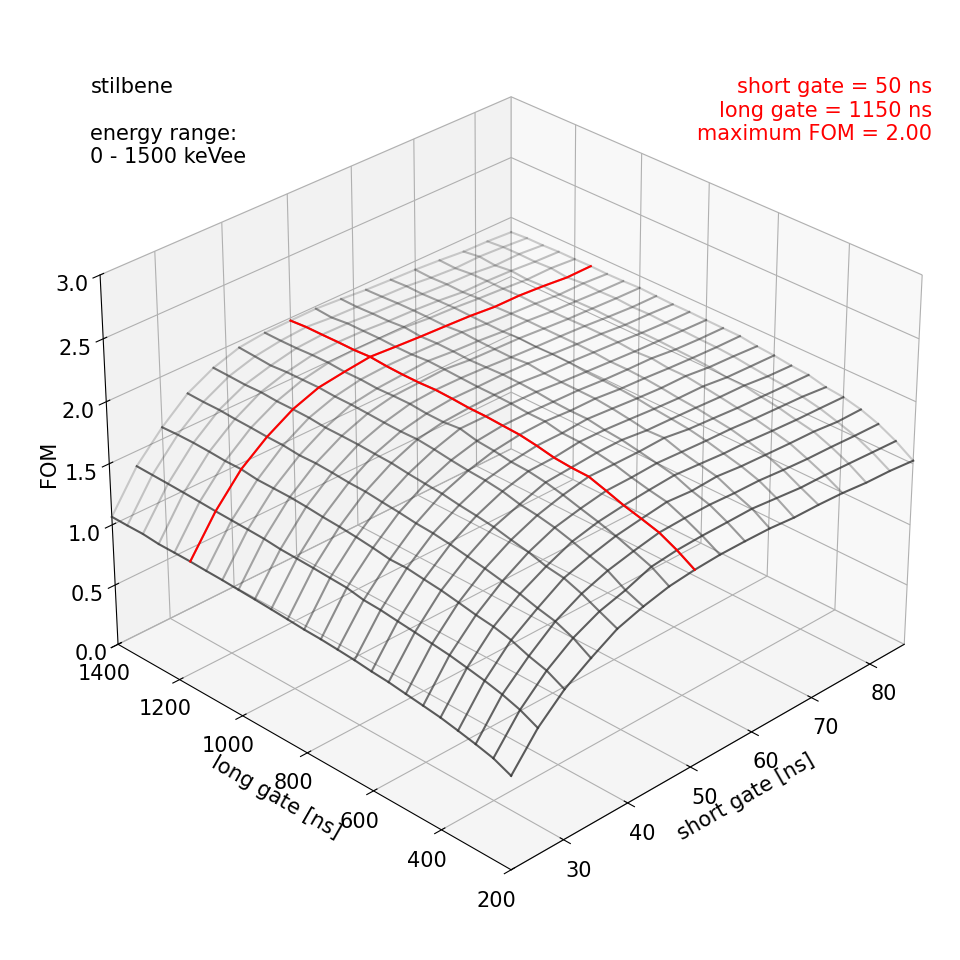}
    \includegraphics[width=0.45\linewidth]{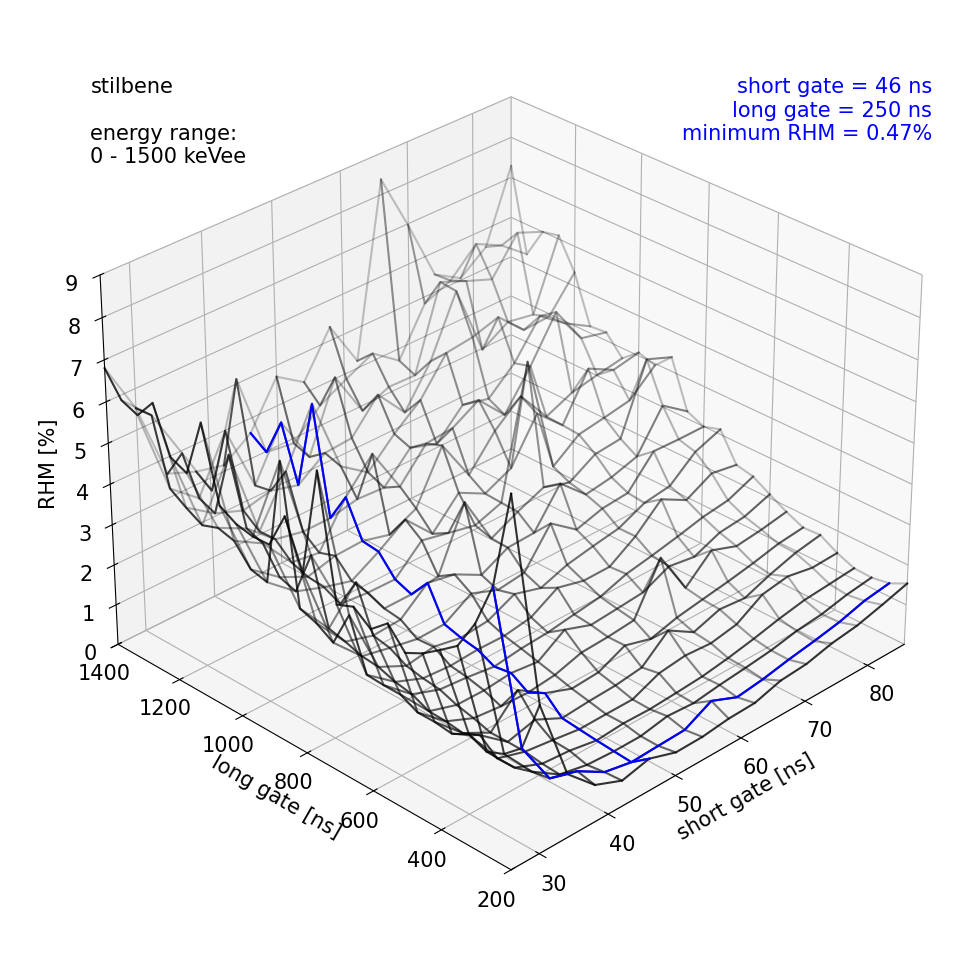}
    \caption{FOM values (left) and RHM (right) for broad energy range at different pairs of gates in case of \mbox{trans-stilbene}. Optimal gates for maximum FOM and minimum RHM are clearly not the same. Long gate was scanned with a step 50~ns and short gate with 4~ns. Gates starting point was set at 16~ns before pulse maximum.}
    \label{fig:min_FOM_stilbene}
\end{figure}

\begin{figure}[H]
    \centering
    \includegraphics[width=0.49\linewidth]{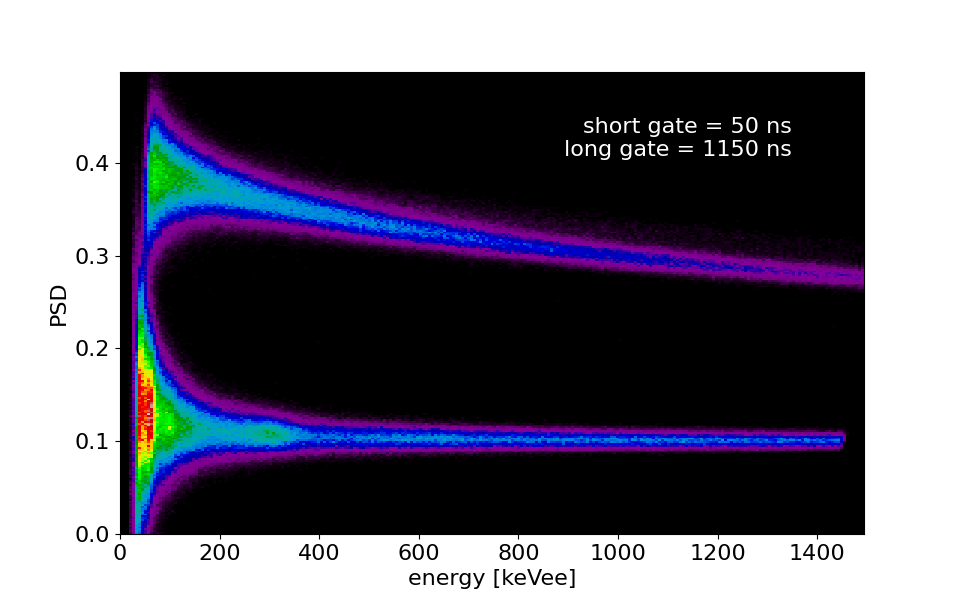}
    \includegraphics[width=0.49\linewidth]{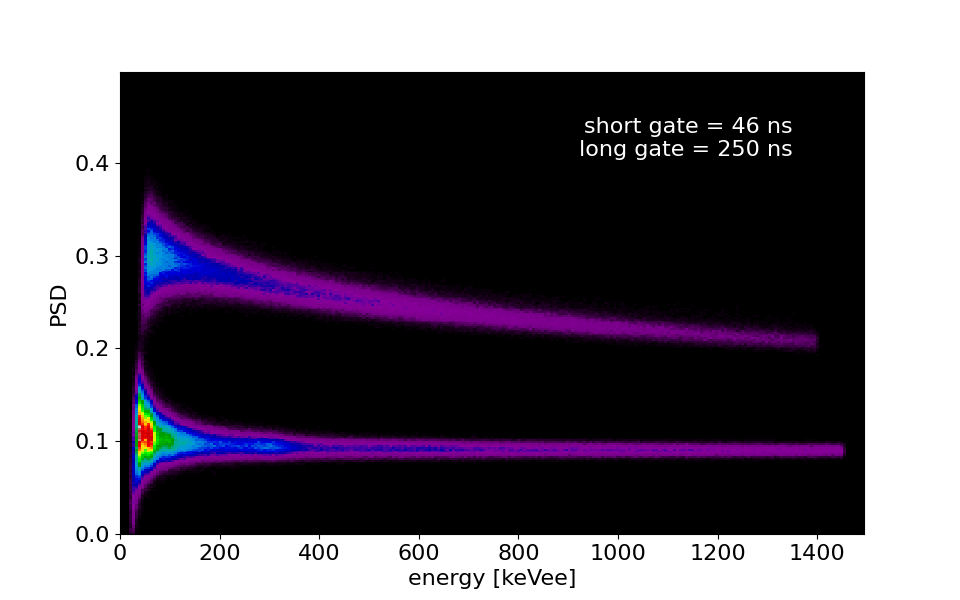}
    \caption{PSD vs energy histograms of \mbox{trans-stilbene} for pairs of gates corresponding maximum FOM (short = 50~ns, long = 1150~ns, left) and corresponding minimum RHM (short = 46~ns, long = 250~ns, right). For minimum RHM, branches of neutron and \mbox{gamma-ray} induced pulses are clearly separated for all energies measured, while for maximum FOM they overlap. Gates starting point was set at 16~ns before pulse maximum.}
    \label{fig:min_hist2D_stilbene}
\end{figure}

FOM value is a convenient way to measure neutron-gamma capability of a scintillator, but it may not be the best representation of all aspects of it. For example, for a broad range energies from 0~keVee\footnote{In off-line analysis we can include all low energy pulses by simply setting 0~keVee as lower energy limit, but of course the real lowest measurable energy is limited by the threshold set in the experimental setup.} to 1500~keVee FOM reaches its maximum value for long gate equal to 1150~ns (see Fig. \ref{fig:min_FOM_stilbene} left). For this gate gamma and neutron branches in 2D heatmap overlap at energy around 50~keVee (Fig. \ref{fig:min_hist2D_stilbene} left).
On the other hand, when looking at PSD vs energy 2D heatmap of \mbox{trans-stilbene} we noticed that for some pairs of short and long gates these branches are clearly separated (see Fig. \ref{fig:min_hist2D_stilbene} right). This is mostly related to the width of PSD peaks, which corresponds to the amount of noise in the long gate. The shorter the gate, the better signal to noise ratio, the better separation. We estimated optimal long gate to separate the branches at 250~ns, which is much less than 1150~ns estimated with best FOM.
\par
\begin{figure}[H]
    \centering
    \includegraphics[width=0.44\linewidth]{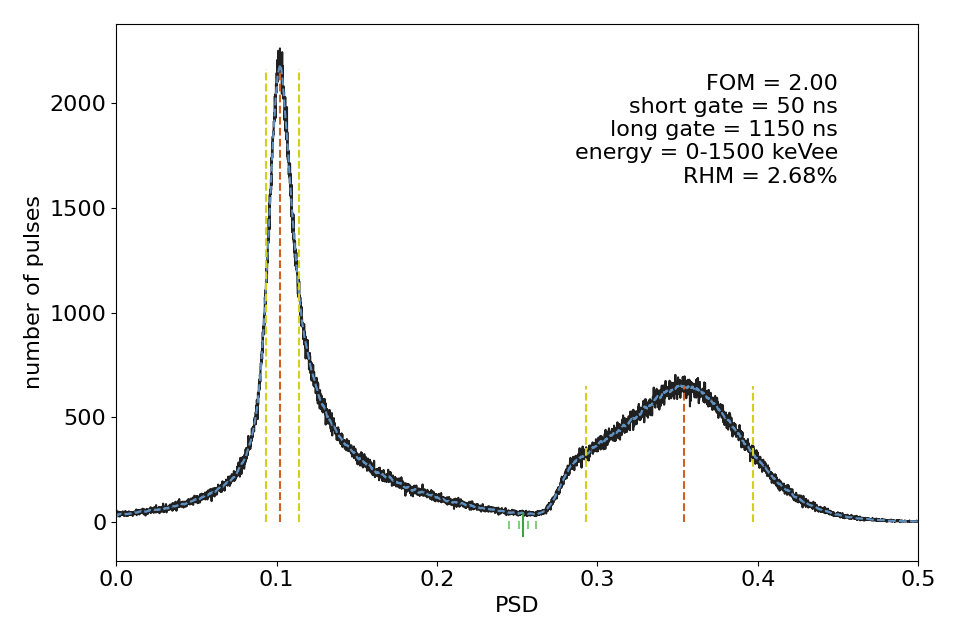}
    \includegraphics[width=0.44\linewidth]{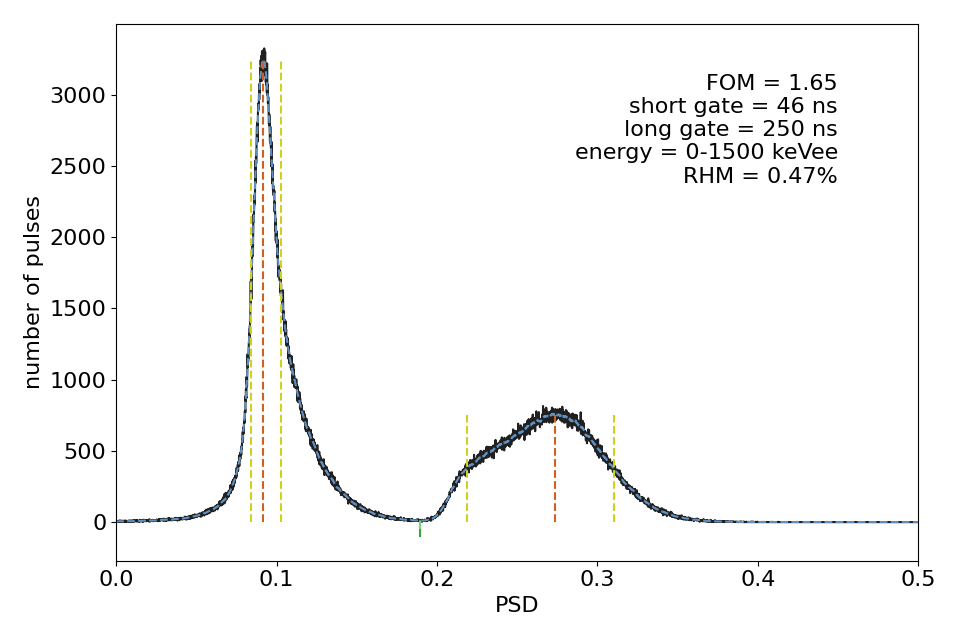}
    \caption{\mbox{Trans-stilbene} PSD histograms for broad energy range for the highest FOM (left) and for the lowest RHM (right). Although FOM is significantly lower in the latter case, the separation of neutron and \mbox{gamma-ray} induced pulses seems to be better.}
    \label{fig:min_PSD_stilbene}
\end{figure}
\begin{figure}[H]
    \centering
    \includegraphics[width=0.49\linewidth]{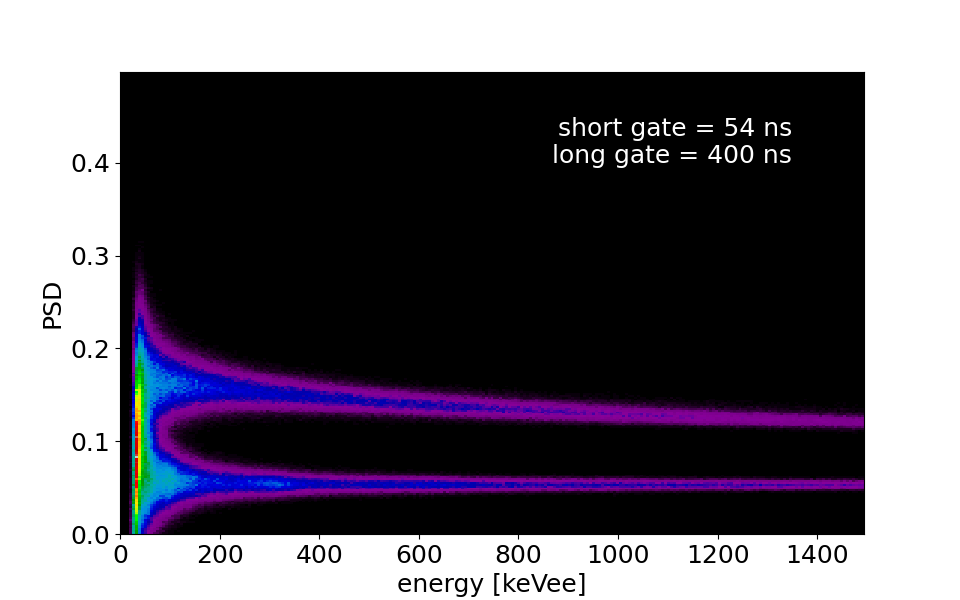}
    \includegraphics[width=0.49\linewidth]{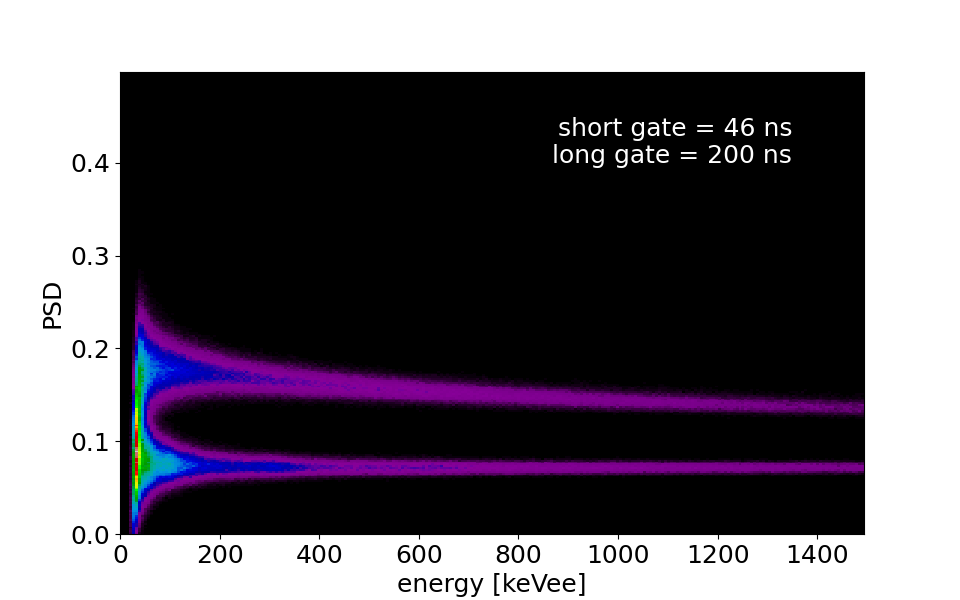}
    \caption{PSD vs energy histograms of OGS for pairs of gates corresponding maximum FOM (short = 54~ns, long = 400~ns, left) and corresponding minimum RHM (short = 46~ns, long = 200~ns, right). Differences are not as clear as in case of \mbox{trans-stilbene}, but again discrimination of neutron and \mbox{gamma-ray} induced pulses at lower energies is better for lower long gate value. Gates starting point was set at 16~ns before pulse maximum.}
    \label{fig:min_hist2D_OGS}
\end{figure}
\begin{figure}[H]
    \centering
    \includegraphics[width=0.44\linewidth]{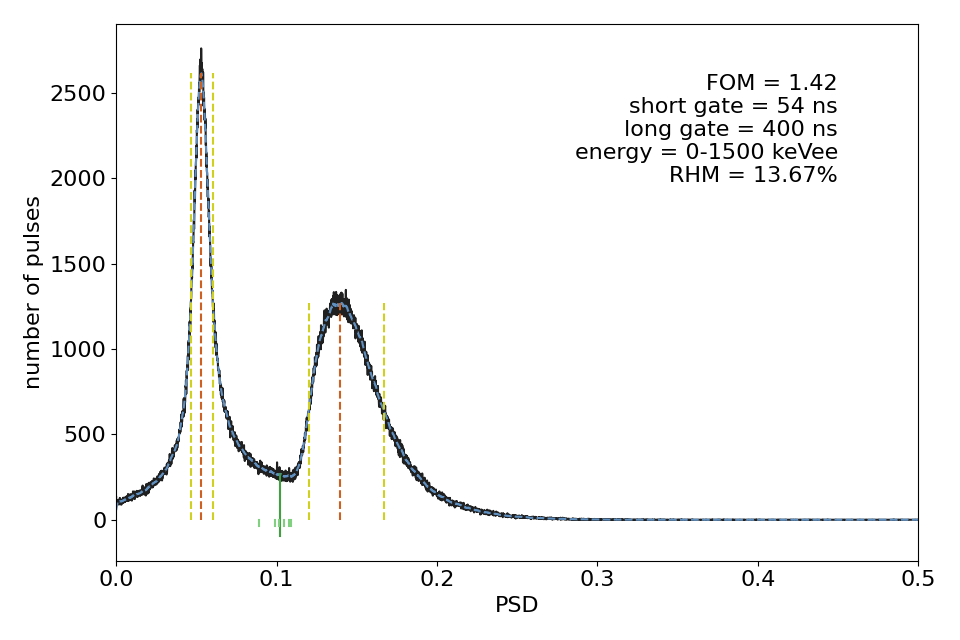}
    \includegraphics[width=0.44\linewidth]{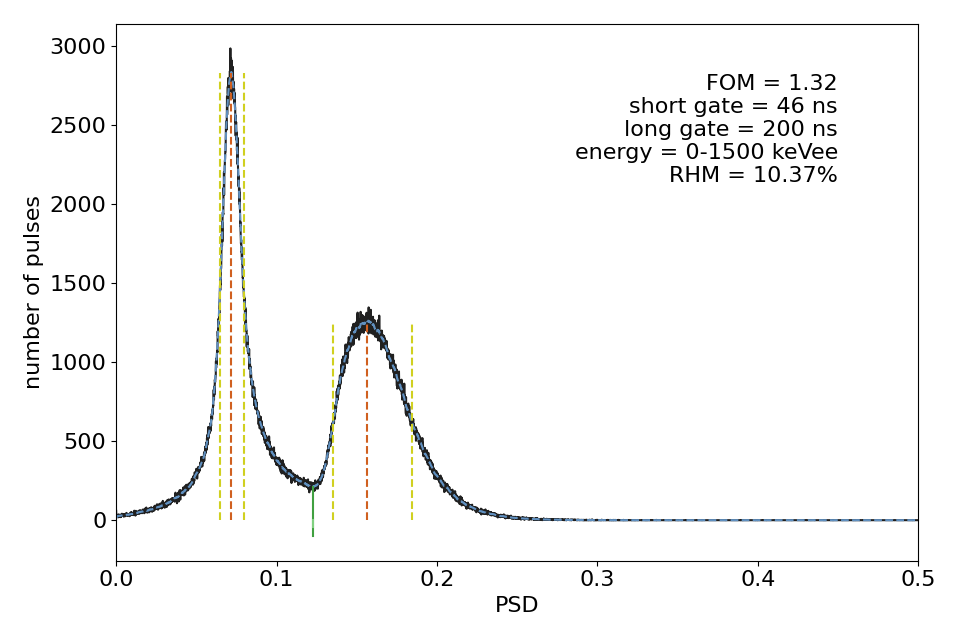}
    \caption{Organic glass scintillator PSD histograms for broad energy range for the highest FOM (left) and for the lowest RHM (right). Differences are not as significant as in case of \mbox{trans-stilbene}, however they still exist.}
    \label{fig:min_PSD_OGS}
\end{figure}

The situation was not so obvious in case of organic glass scintillator, since there is no full separation of branches. They always overlap at low energies (see Fig. \ref{fig:min_hist2D_OGS}).
Yet still we observed that for some pairs of gates the overlapping part is smaller and the minimum between the peaks in PSD histogram is lower (see Fig. \ref{fig:min_PSD_stilbene} and \ref{fig:min_PSD_OGS}). 
As this height of minimum changes along with the peaks, we decided to express it relatively to the average of peaks heights:
\begin{equation}
    \label{eq:min}
    RHM [\%] = \frac{2 h_{min}}{h_{n}+h_{\gamma}}
\end{equation}
where $RHM$ stands for Relative Height of Minimum, $h_{n}$ is the height of neutron peak, $h_{\gamma}$ is the height of gamma peak and $h_{min}$ is the height of the minimum above the zero axis.
The lower RHM is, the better the separation.
\par
\begin{figure}[H]
    \centering
    \includegraphics[width=0.45\linewidth]{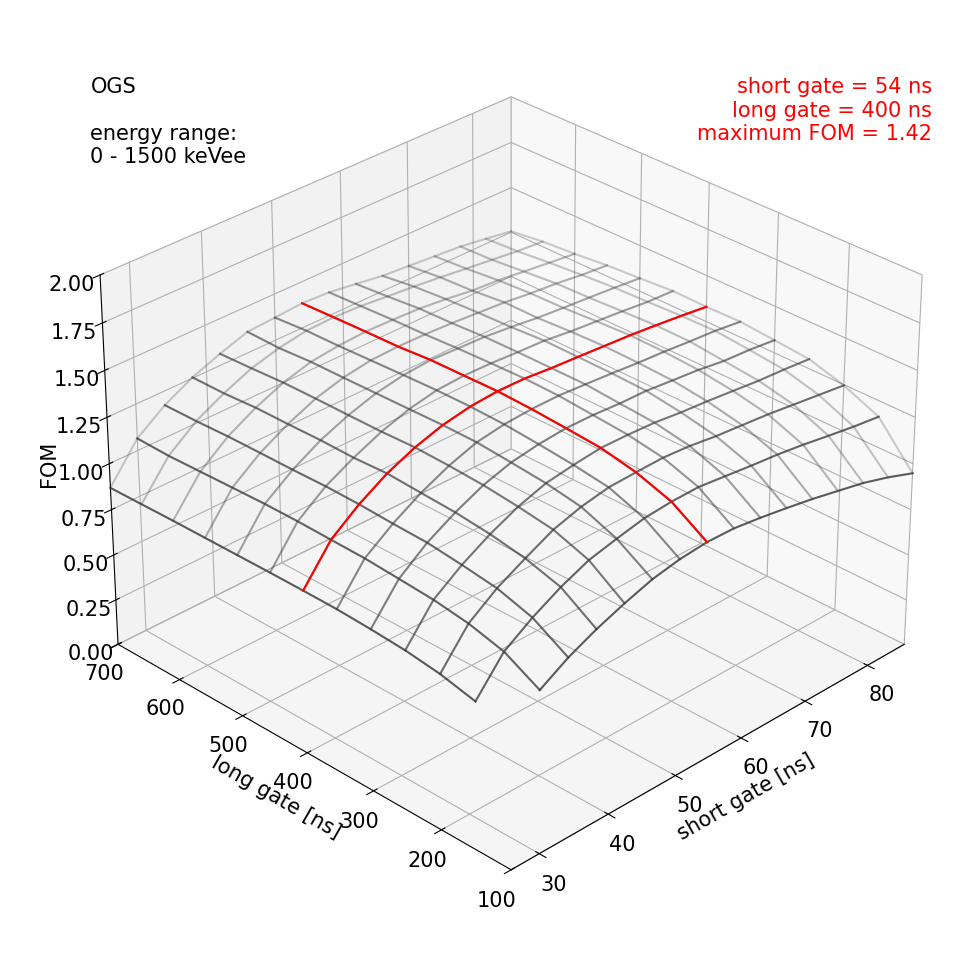}
    \includegraphics[width=0.45\linewidth]{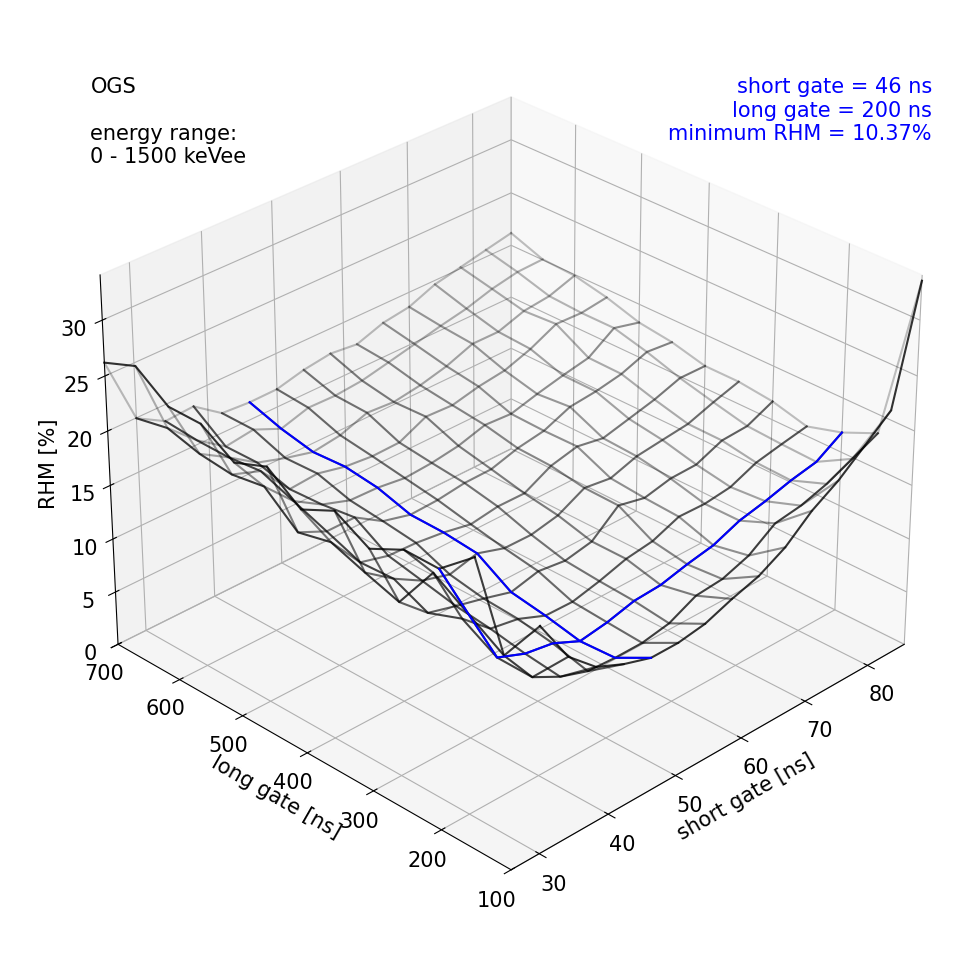}
    \caption{FOM values (left) and RHM (right) for broad energy range at different pairs of gates in case of OGS. Optimal gates for maximum FOM and minimum RHM are not the same. Long gate was scanned with a step 50~ns and short gate with 4~ns. Gates starting point was set at 16~ns before pulse maximum.}
    \label{fig:min_FOM_OGS}
\end{figure}

As we can see in Figures \ref{fig:min_FOM_stilbene} and \ref{fig:min_FOM_OGS}, the minimal RHM and maximum FOM values are indeed obtained for different gates.
The difference is much more subtle in case of OGS than it is for \mbox{trans-stilbene}, but nonetheless, it is significant. Unfortunately, our RHM finding procedure was not perfect and sometimes gave only estimated value, thus RHM graphs in Figures \ref{fig:min_FOM_stilbene} and \ref{fig:min_FOM_OGS} have significant amount of noise, but the overall tendency is clear. We hope to improve this procedure in the future.

This observation brings some questions.
\par
Is Figure of Merit a good indicator of \mbox{neutron-gamma} discrimination capability?
We believe it is, as in a range of values between 0.5 and 2.0 it does indeed represent this feature of scintillator accurately. On the other hand, all values above 2.0 mean almost perfect separation of neutron and gamma pulses, so in practical applications there is no difference between scintillators which have FOM higher than 2.0. Another issue with Figure of Merit is that it is well defined only in relatively narrow energy ranges, where PSD peaks are \mbox{Gaussian-like}. It is much harder to calculate FOM reliably for broad energy range, and therefore it is hard to compare scintillators that have different energy ranges of applicability.
Of course, higher FOM values in a narrow energy range may imply better separation in broader energy range, but it is not always true. Another complementary indicator would be useful and Relative Height of Minimum may be a good candidate.
Similar minimum proved to be useful as an discriminator in a software for automated estimation of neutron/photon misclassification developed in University of Michigan \cite{Polack2015253}.
\par
What gates should be chosen for measurements?
It depends on the goal of measurement. In scientific research the procedure of choosing gates should guarantee that the results are reliable and repeatable. FOM is well established in this role and its maximum value seems to be good point of reference. However, it must be said what energy range was used to estimate the gates to measure FOM, because they are energy dependent (see Section \ref{sec:gates}). On the other hand, in practical application of neutron detection it would be better to search for gates that allow to separate neutrons from \mbox{gamma-rays} in the broadest possible energy range. They can be estimated by looking at PSD vs energy 2D histograms or alternatively RHM may be a good numerical indicator.

\section{Conclusions} \label{sec:conclusions}

Charge Comparison Method is used in many scientific research involving Pulse Shape Discrimination and the results are often reported as FOM values for given pairs of short and long gates. As we have shown, gates starting point (rarely reported) is also very important, because it can have significant impact on calculated FOM. It is even possible to improve FOM values by deliberately starting gates near pulse maximum. This can be especially beneficial in a hardware with low bandwidth.
\par
Optimal short and long gates are energy dependent. The reason for this is unclear to us and we are going to investigate it further, but nonetheless the conclusion is that it is important to apply the same procedure of finding optimal gates to make comparison of scintillators accurate.
\par
Low energy neutron and gamma pulses can be separated better for different pair of gates than suggested by maximum FOM value. Therefore, it would be useful to define another value that indicates the quality of separation.
Relative height of minimum between peaks in PSD histogram is a simple candidate. However, it has to be tested fully to determine its scope of applicability.
\par
In practical applications of neutron detection over a large range of energy it is beneficial to manually tune the gates and the starting point to obtain the best neutron-gamma discrimination, regardless of whether they cover whole of the measured pulse. This seems counter-intuitive, as some of charge information may be lost, but we have seen improvements in our results even when most of the pulse shape was ignored.
\par
Gates chosen with RHM are shorter than these chosen with FOM. This makes sampling window of the signal potentially shorter and input rate potentially higher.
\par
Both FOM and RHM are affected by electrical noise in the signal. Unfortunately, it is hard to reduce the effects of this noise in \mbox{post-processing}. Therefore, it is advisable to keep noise level as low as possible during the measurements to obtain reliable results.
\par
In case of \mbox{neutron-gamma} discrimination Constant Fraction Discrimination triggering method is more reliable than Leading Edge. Both can be replaced by triggering at pulse maximum, which is equivalent to CFD method, but simpler. It can be applied in \mbox{off-line} analysis of digitally recorded pulses and is easy to implement in the hardware.
\par

\section*{Acknowledgements}
This work was supported in part by the European Union (ChETEC-INFRA, project no. 101008324).

\bibliographystyle{unsrtnat}
\bibliography{bibliography}

\end{document}